%% file: main.tex
\documentclass[11pt, letterpaper]{article}

\usepackage{booktabs,placeins,hyperref,todonotes,enumitem,longtable,array}
\usepackage{natbib,microtype,mathptmx,setspace,array,smartdiagram,pdfpages}
\usepackage{authblk}
\usetikzlibrary{arrows.meta,arrows,shapes,positioning,shadows,trees}
\tikzset{
  basic/.style  = {draw, text width=2cm, drop shadow, rectangle},
  root/.style   = {basic, rounded corners=2pt, thin, align=center,
                   fill=blue!20},
  level 2/.style = {basic, rounded corners=6pt, thin,align=center, fill=orange!50,
                   text width=9em},
  level 3/.style = {basic, rounded corners=6pt, thin, align=left, fill=pink!50, text width=7em}
}

\usepackage[no-weekday]{eukdate}
\usepackage[margin=1.9cm]{geometry}

\setlist[description]{leftmargin=.5cm,itemsep=0.2ex,font=\itshape}

\newcolumntype{A}{>{\raggedright\arraybackslash}p{3.4cm}}
\newcolumntype{B}{>{\raggedright\arraybackslash}p{3.4cm}}
\newcolumntype{C}{>{\raggedright\arraybackslash}p{10.1cm}}

\usepackage{caption}
\DeclareCaptionStyle{italic}[justification=centering]
 {labelfont={bf},textfont={it},labelsep=colon}
\usepackage{graphicx}
\title{Good intentions, unintended consequences: exploring forecasting harms}

\author[1*]{Bahman Rostami-Tabar}
\author[2]{Travis Greene}
\author[3]{Galit Shmueli}
\author[4]{Rob J. Hyndman}

\affil[1]{Data Lab for Social Good Research Group, Cardiff Business School, Cardiff University, UK}
\affil[2]{Department of Digitalization, Copenhagen Business School, Denmark}
\affil[3]{Institute of Service Science, National Tsing Hua University, Taiwan}
\affil[4]{Department of Econometrics \& Business Statistics, Monash University, Australia}

\begin{document}
\maketitle

\begin{abstract}
	Organizations worldwide that rely on data-driven approaches regularly employ forecasting methods to enhance their planning and decision-making processes. While extensive research has examined the harms associated with traditional machine learning applications, relatively little attention has been given to the ethical implications of time series forecasting. However, forecasting presents distinct ethical challenges due to its diverse organizational applications, varied objectives, and unique data processing, model development, and evaluation workflows. These distinctions complicate the direct application of existing machine learning harm taxonomies to common forecasting scenarios. To address this gap, we conduct multiple interviews with industry experts and academic researchers, systematically identifying and analyzing under-explored domains, use cases, and potential risks associated with forecasting. Our objective is to develop a novel taxonomy of forecasting-specific harms. Drawing inspiration from Microsoft Azure’s taxonomy for responsible innovation, we integrate a human-led inductive coding approach with AI-driven analysis to extract key categories of harm in forecasting. This taxonomy aims to support researchers and practitioners by fostering ethical reflection on their decision-making throughout the forecasting process. Additionally, we seek to establish a research agenda focused on identifying measures to mitigate potential harms in forecasting. By highlighting unique risks within forecasting, our work contributes to the broader discourse on machine learning ethics.
\end{abstract}

\section{Introduction}

Time series forecasting is, broadly, the process of using historical data points collected sequentially over time to predict future values. Modern forecasting techniques enable organizations to more effectively plan, manage risk and deal with uncertainty by estimating likely future outcomes. Forecasting models allow organizations to avoid problems that may arise by simply repeating previously successful actions from the recent past, imitating the actions of successful peers, or else following their leaders' often-questionable ``gut instincts''.



Despite the pervasive use and benefits of forecasting in business and society, we know surprisingly little about the harms specific to forecasting. Even popular textbooks on forecasting \citep{montgomeryTimeSeriesBook2015, hyndman2018forecasting} do not substantively engage with the issue of harm\footnote{The single mention of ``harm'' we found comes from the 600-plus-page textbook \citet[pg. 21]{montgomeryTimeSeriesBook2015}, who asks the reader to discuss ``the harm or penalty associated with a bad forecast'' as a homework exercise.} or discuss responsible forecasting practices. Very little work exists on these topics, yet in our experience most forecasters see value in being more transparent about its potential harms on individuals and society. This apparent tension piqued our curiosity.


 Indeed, our interest in the topic of  forecasting harm gradually emerged from candid discussions with forecasting experts in both industry and academia.  Forecasters frequently lamented the lack of awareness and the need for a structured analysis of forecasting’s potential harm. For instance, the Editor-in-Chief of Foresight at the International Institute of Forecasters described the topic of forecasting harm as an ``unexplored area,'' stating, ``I hadn't heard this question posed like this before, but just bringing awareness makes a difference.'' He emphasized that identifying common ways forecasting introduces harm could itself help mitigate negative effects. Further insights emerged from a panel on ethical considerations in forecasting organised by the lead author, featuring experts from the International Organization for Migration, UNHCR Innovation, the Neukom Institute for Computational Science, and Yale University. Panellists discussed forecasting’s unintended harms in migration, refugee crises, and public health, emphasizing the need for research and guidelines to mitigate negative consequences. They underscored how forecasting errors can have real-world impacts, affecting vulnerable populations and leading to resource misallocation. Industry case studies further reinforced these concerns around the social impact of forecasting---a Google data scientist highlighted how flawed forecasts contributed to Target Canada’s collapse, resulting in a \$2.1 billion loss and the closure of 133 stores. Similar concerns were echoed in our interviews of 21 experts from academia and industry. One respondent noted, ``I’ve seen people so fascinated by producing forecasts that they don’t necessarily think about where it will be used, either well or poorly, for good or for harm” (P21). Another participant emphasized the importance of creating awareness: ``Rather than accepting a set of signals, we should pause and think about how we consume this information” (P18). Overall, our discussions with forecasting experts in industry and academic shaped our research agenda by highlighting a few key gaps: the lack of a systematic treatment of forecasting harms, the tangible consequences of using forecasts across domains, and the need for ethical guidelines to mitigate the risks of harm.

What might explain these gaps in our understanding of forecasting harms? We offer several tentative hypotheses. First, forecasting often takes place behind closed doors, making it difficult for researchers external to an organization to study. Second, forecasting---in contrast to traditional supervised learning with cross-sectional data---generally focuses on predicting the behavior of aggregates rather than individual units, making it hard to identify harms to specific individuals or organizations. When harms to individuals are thinly spread out over society, no one single person is motivated enough to speak out. Also, the focus on aggregates also means that personal data generally plays a smaller role in forecasting, which may contribute to less legal and regulatory scrutiny. Third, unlike some applications of AI and machine learning that have provoked major media and journalism reporting (e.g., social credit scoring, facial recognition, and criminal risk predictions), applications of forecasting have tended to receive much less popular coverage. This last reason is perhaps again because much information about forecasting stays within organizations, making it difficult for an external individual or group to contest the quality or accuracy of a forecast.

Three objectives therefore motivate this study. 1) \emph{What are the harms specific to forecasting?}; 2) \emph{How might such harms be mitigated?} and 3) \emph{What should a research agenda for responsible forecasting entail?} Answers to these questions lead to a taxonomy of forecasting-specific harms, and a practical toolbox of harm mitigation strategies. Ultimately, we hope our research can contribute to understanding and awareness about where and how forecasts can be used for good, and promote ethical reflection and critical discussion on the role of forecasters in society.


 Note that for the purpose of this study we distinguish forecasting from other similar terms, such as ``predictive AI'' or ``predictive analytics,''\footnote{In practice many of the same statistical forecasting methods are used atemporally to, for instance, classify objects, estimate past numeric values, or retrieve information  \citep{barocas-hardt-narayanan}.} using the following heuristic criteria:
1) forecasting is forward-looking, providing estimations of the variable of interest's value at one or more future time periods;
2) forecasting is based on historical observations of the variable of interest (i.e., time series);
and
3) the forecasting models can include past and future values of other exogenous variables \citep{hewamalage2021recurrent,  makridakis2020social}.

This paper is structured as follows. \autoref{sec:literature} briefly surveys relevant legal and philosophical literature to examine the concept of harm with the goal of relating these considerations to issues in forecasting. \autoref{sec:method} explains our methodology based on a combination of inductive coding and AI-assisted categorization. Our results and typology are in \autoref{sec:typology}. \autoref{sec:discussion} discusses our results, and \autoref{sec:FutureDirections} proposes future research directions for the ethics of responsible forecasting. We conclude with a summary of our findings and some final comments in \autoref{sec:conclusions}.

\section{Harms in forecasting}\label{sec:literature}

\subsection{The concept of harm}

Recognizing a forecasting harm requires a theory of harm to guide questions of moral and legal responsibility for harms. Yet, as with many concepts that have ethical and political implications, the concept of harm is interrelated with contested notions of \emph{wrongs, justice, duties, rights, responsibility, interests, intentions, causality, care, liability,} and \emph{consent}, to name just a few \citep{feinberg1984harm}. This makes it hard to define harm without begging the question. Our goal is not to articulate a novel theory of harm specific to forecasting, but to leverage the pre-existing harm-related concepts and frameworks developed and refined by legal scholars, philosophers, and (increasingly) AI ethics researchers.

Nearly all human cultures have subscribed to some version of the idea that we should, where reasonable, avoid doing harm to others \citep{bradley2012doing}. The prohibition on harm is so important that it is presumed to also apply to intelligent machines created by humans. For instance, the first of Isaac Asimov's influential ``three laws of intelligent machines'' begins with the command: ``A robot may not injure a human being, or, through inaction, allow a human being to come to harm'' \citep{asimov1978bicentennial}. So while there is nearly universal consensus that intelligent machines should avoid inflicting harm on humans, a more difficult question is what constitutes a harm. Here we turn to fields that specialize in examining definitional questions: philosophy and law.

Philosophical and legal accounts generally focus on three senses of harm:  1) something being damaged or broken; 2) unjust or wrongful treatment; and 3) the ``defeating of an interest'' \citep[pg. 33]{feinberg1984harm}. In this initial study, we treat the concept of harm as a cluster concept, and provisionally define harm as \emph{unjustly or wrongly setting back an interest of an individual or collective of individuals} (e.g., an organization).\footnote{It is worth noting---especially since the AI ethics literature often invokes human rights \citep{aizenberg2020designing}---that rights function as legal devices to protect an entity's interests and imply (sometimes burdensome) correlative duties on others \citep{onora2002autonomy}. Legal documents that draw on fundamental rights, such as the EU's GDPR, commonly rely on ``balancing tests'' to consider how the exercise of a data subject's right justifies imposing a correlative duty on another entity (i.e., an organization wishing to process such data for business reasons)  \citep{greene2019adjusting}.} Future research might explore how to best define the normative terms ``unjust'' or ``wrong'' in the context of forecasting.

\subsection{Typologies of AI harms}
Diverse communities of researchers with interests in ``AI for social good'' \citep{goodsocietyfloridi2021ethical} and Fairness, Accountability, Transparency, and Ethics (FATE) in machine learning \citep{wieringa2020account}, have singled out a number of harms stemming from the use of supervised, unsupervised and reinforcement learning algorithms. These harms include the maintenance of historically unjust power relations, the generation of toxic language, the promotion of addictive behaviors, and the fostering of various informational harms related to, for instance, the spread of polarizing news or conspiratorial beliefs in society \citep{chan2023harms}. AI ethics researchers have also begun to study the dynamics and harms of self-fulfilling prophecies when AI/ML models are applied in social contexts \citep{king2023self}. 

Technological harms can arise at multiple levels of society. \citet{shelby2023sociotechnical} constructed a taxonomy of sociotechnical harms at micro-, meso- and macro-levels stemming from the use of algorithms. They thematically list the harms as representational harms, allocative harms, quality of service harms, interpersonal harms, and social system harms. Broadly, these harms relate to how social groups are represented in inputs and outputs, how these representations influence resource distribution decisions, how optimization choices relate to model performance for various users, how the model outputs affect social relations 
and properties of social systems, such as their level of stability or equality. 
Due to forecasting's traditional focus on aggregate behavior, allocative and social systems harms appear the most relevant to forecasting, while representational harms (i.e., the reinforcement of undesirable social and cultural stereotypes and beliefs) seem less relevant (see e.g., \citet{barocas2017problem, suresh2021framework}).

As with many new technologies that are rapidly introduced into society, newly identified harms can ignite streams of research. For instance, the rise of deep neural networks and large labeled image datasets such as ImageNet, stimulated a wave of research into how various morally-charged labels were associated with certain demographic groups \citep{denton2021genealogy}. The rise of large language models (LLMs) has also led to a growing body of research focused on harms specific to LLMs. There are documented discrimination, exclusion and toxicity, and human-computer interaction harms, as well as automation, access, and environmental harms \citep{weidinger2022taxonomy}.

Industry researchers are the often the first to study the harms from new AI technologies. A white paper from Microsoft, which we refer to as the Microsoft Azure framework \citep{msharms}, specifically addresses responsible innovation and technology harm mitigation strategies. The framework classifies technology harms into four types: risk of injury, denial of consequential services, infringement on human rights, and the erosion of social and democratic structures. \emph{Risk of injury} refers to physical injury due to, for example, over-reliance on safety features, inadequate fail-safes, and exposure to unhealthy agents during technology manufacturing or disposal. It also covers emotional or psychological injury due to emotional distress, distortion of reality, reduced self-esteem, addiction, identity theft, and misattribution. \emph{Denial of consequential services} reflects opportunity loss such as employment, housing, insurance, and educational discrimination, as well as digital divide and loss of choice. Technologies that automate decisions can lead to credit discrimination, differential pricing, economic exploitation, and devaluation of individual expertise. \emph{Infringement of human rights} involves privacy, liberty, and dignity harms via dehumanization and changes in how people perceive and engage with each other in society. The \emph{erosion of social and democratic structures} includes harms related to amplification of power inequality, behavioral manipulation and deception, and stereotype reinforcement.

\subsection{From harms in AI to harms in forecasting}
\label{sec:forecastingprocess}

While research on responsible AI and machine learning is growing \citep{dignum2019responsible, barocas-hardt-narayanan}, translating and applying these ideas and methods to forecasting pipelines is not straightforward. There is a pressing need for such work given the scale of forecasting's influence on human behavior and social systems.
One notable exception is \citet{hobday2019ethical}, who examines ethical issues stemming from forecasting in the context of marine ecology, and proposes ten guiding principles for ethical forecasting. He recommends transparency, accurate uncertainty representation, stakeholder education, and performance reviews. These principles, although drawn from the specific domain of marine ecology, broadly align with the harm mitigation strategies we describe in Section \ref{sec:harmmitigation}.

The lack of more general work on forecasting harms is surprising for several reasons. First, many forecasting applications involve aspects of automated decision-making. For example, in retail and e-commerce, items are automatically ordered based on forecasted demand of ``replenishables''. AI and data protection regulations aimed at reducing the harms of ``high risk'' automated systems, usually subject automated systems to greater regulatory scrutiny than systems where humans are ``in the loop'' \citep{binns2022human}. Second, forecasting can involve system-wide scales with catastrophic stakes, as when forecasting events such as financial crises, terrorist attacks, earthquakes, and floods \citep{sornette2002predictability}. Third, societies around the world are witnessing rapid growth in the number of sensors and devices that capture and analyse individual-level time series data. Thus, forecasting has the potential to impact individuals, societies, and ecosystems as forecasts feed into increasingly complex automated decision-making systems.

These factors suggest the need for systematic study of the harms of forecasting and possible mitigation strategies aimed at promoting responsible forecasting practices. A better understanding of the harms of forecasting can lead to more socially aligned applications of forecasting \citep{thompson2022escape}. As one recent popular work on the ethics and politics of AI notes, ``[the] separation of ethical questions away from the technical reflects a wider problem in the field, where the responsibility for harm is either not recognized or seen as beyond the scope of the research'' \citep{crawford2021atlas}. Related work in psychology and AI ethics also highlights how AI tools can inflate our ``moral distance'' to the needs and interests of others remote in time and space \citep{villegas2023moral}, contributing to an undesirable form of ``moral disengagement'' \citep{bandura2002selective}, and potentially blinding us to how our forecasting activities impact the interests and well-being of others. Our work attempts to address these gaps surrounding the ethics of forecasting by focusing on identifying various types of harm that may occur in the context of forecasting. The identification of harms can promote more reflexive and responsible forms of forecasting. 

Lastly, a simplifying assumption guides our analysis of the forecasting process. We suppose that, regardless of sector or decision, the forecasting workflow is relatively stable across forecast types and applications. Thus, the typologies of harm and the mechanisms by which they may occur can be generalized, identified, and potentially mitigated when the workflow is described in a standardized way. 

\subsection{Harms of forecasting: preliminary ethical and legal considerations}
The study of forecasting harm is still in its infancy. As a result, in many real-world cases where forecasts seem to cause harm, considerable disagreement pervades about who or what was harmed (and how), as well as who is responsible. To make progress on such important and practical issues, we suggest that an adequate theory of forecasting harm should provide reasonably clear answers to a basic set of questions. We therefore offer some preliminary theoretical desiderata, drawn from the fields of ethics and law. While not exhaustive, these initial desiderata can serve as normative benchmarks for the analysis and discussion of forecasting harms as understood by practising forecasters (Section \ref{sec:typology}).  An overview of these theoretical considerations and relevant questions are given in \autoref{tab:theory_harm}.

\begin{table}[!htbp]
	\caption{Considerations for a theory of harm from law and philosophy, and their relevance to the harms of forecasts.}
	\label{tab:theory_harm}
	\centering
	\begin{tabular}{%
		>{\raggedright\arraybackslash}p{4.5cm}%
		>{\raggedright\arraybackslash}p{5.4cm}%
		>{\raggedright\arraybackslash}p{6.4cm}}
		\toprule
		\textbf{Theory of harm desiderata}
		 & \textbf{Description}
		 & \textbf{Relevance to forecasting}
		\\
		\midrule
		Individual vs. Collective Harm
		 & What kinds of entities have interests?
		 & Do social and ecological systems have interests that can be unjustly defeated by forecasting?
		\\
		\midrule
		Comparative vs. Absolute Harm
		 & Must an account of harm consider how things could have gone?
		 & What are the likely differences in potential harmful outcomes when publishing and not publishing a forecast?
		\\
		\midrule
		Harm vs. Imposition of Risk
		 & Must a harmful act result in observable consequences?
		 & How should we think about poorly designed forecasting procedures that impose unjustifiable risk on society and individuals despite not realizing harmful consequences?
		\\
		\midrule
		Intentions and Unforeseen Consequences
		 & Does moral blame for harm depend on whether such harms were indirect or directly foreseen or intended?
		 & To what extent can forecasters foresee an individual or collective reaction to the forecast?
		\\
		\midrule
		Standards of Care and Negligence
		 & To what extent is an activity inherently dangerous and risky, thereby imposing on participants a duty of care and liability for harm?
		 & In what domains does forecasting constitute a dangerous or high-risk activity? What is the professional relation of forecasters to individuals and society?
		\\
		\bottomrule
	\end{tabular}
\end{table}


\begin{description}[leftmargin=.5cm,itemsep=0.2ex,font=\itshape]
	\item[Individual vs collective harms] We presume that collectives such as human societies, and non-human collectives such as ecosystems, have interests that can be thwarted (see e.g., \citet{taylor1986respect}). We treat social and collective harms as synonyms, while noting some legal scholars argue that \emph{societal harm} is conceptually distinct from individual and collective harm \citep{smuha2021beyond}.

	\item[Comparative vs absolute harms] Comparative theories of harm draw on comparisons of actual well-being to counterfactual well-being had the event not occurred \citep{klocksiem2012defense,bradley2012doing}. In contrast, absolute harms do not take into account how things could have happened (but did not); they often involve causing an entity to enter an intrinsically bad state, such as pain, mental or physical discomfort, disease, deformity, disability, or even death \citep{harman2009harming}. Related to the idea of absolute harm is the view that the mere violation of a right (i.e., a protected interest) constitutes a harm \citep{feinberg1984harm}. Clarifying whether a claim of forecasting harm implicitly draws on a comparative or absolute theory of harm can help in determining whether, for instance, a group of individuals was truly harmed from the decision to publish a forecast.\footnote{The issue of euthanasia helps illustrate key differences between comparative and absolute theories of harm. In absolute accounts of harm, death is usually presumed to be a bad state, and so euthanasia constitutes a harm. But in the comparative account of harm, whether euthanasia is harmful depends on whether death is preferable to (otherwise) prolonged suffering. Under the comparative account of harm, depending on one's prognosis, euthanasia can receive a positive moral evaluation.}

	\item[Harm vs imposition of risk] A further distinction relevant to forecasting is how and whether to separate harm from the imposition of risk. Distinguishing the two is not always simple, making an ethical analysis of forecasting complicated. This is because not all harms materialize \citep{finkelstein2002risk}, as when a drunk driver manages to safely drive home at night. Legal scholars and philosophers debate whether the imposition of risk is itself a harm (see e.g., \citet{hayenhjelm2012moral}). To simplify our analysis, we treat them as interchangeable for now.

	      A related legal and philosophical concept is that of \emph{moral luck} \citep{nagel1979mortal}, which can arise when we blame an agent for some harm whose realization is stochastic in nature. Moral luck captures the intuition that moral judgments of agents can depend on factors outside of their control. For instance, although eventually most of the convictions were overturned on appeal, the scientists punished in the L'Aquila earthquake incident arguably experienced moral (bad) luck, at least initially. The phenomenon of moral luck suggests a  satisfactory theory of harm and punishment should not only consider actual outcomes, but also an agent's intent \citep{cane2002responsibility}.

	\item[Intentions and (un)foreseen consequences] Ethical theories, particularly those drawing on Kantian or virtue-based conceptions of ethics, emphasize the motives and intentions of agents as relevant when evaluating responsibility for harm. Related to the issue of intent to harm is the foreseeability of harm. Consider a forecaster's ability to foresee how others may react to the disclosure of forecasts. Reactions to forecasts can themselves cause harm, sometimes systemic harms, as in bank runs whose effects can spell ruin for entire banking systems \citep{rochet2009there}. Presumably, however, the forecaster does not intend to cause a bank run.


	      This example indicates an ethical analysis of harms related to forecasting may therefore distinguish between harms that are ``actively'' intended, from those merely ``allowed'' to happen indirectly as a byproduct of the disclosure. Here the \emph{doctrine of double effect} \citep{foot1967problem,quinn1989actions} seems relevant. The doctrine implies that forecasters are less morally responsible for the effects of their actions (i.e., disclosure) done knowingly but unintentionally.

	\item[Standards of care and negligence] Civil or ``tort'' law concerns wrongs done by individuals to other individuals in society, often out of neglect \citep{cane2002responsibility}. Tort law concepts can clarify the nature of the relationship between forecasters and society, and help decide whether societal stakeholders are owed a duty of care. If this duty of care can be shown to have been breached, then forecasting organizations may be found liable for any resulting harm. But as the L'Aquila incident makes clear, causal responsibility for harm does not always imply legal or moral responsibility \citep{smiley1992moral}. Depending on the level of risk inherent to the activity, legal responsibility arises when persons and organizations fail to exercise due care while performing the activity. Setting standards of reasonable or ``due'' care encourages people and organizations to take appropriate precautions to avoid harming others, often at a cost. Legal systems seek to discern the appropriate---or just---balance between the reduction in harm and costs of precaution. Altogether, the issues of due care and negligence raise questions of the stakes of forecasting and what constitutes reasonable precaution when performing the activity of forecasting. Professionalization of forecasting, along the lines of medicine or law, could provide the institutional means of providing concrete answers to these harm-related questions, but likely at the cost of a reduction in forecasters' freedom. 
\end{description}

\section{Methods}
\label{sec:method}

\subsection{Data collection and analysis techniques}

In this paper, we used a combined inductive-deductive research approach. We began our investigation with a deductive analysis, focusing on the categories of harms defined in the previously-mentioned Microsoft Azure framework, along with the general theory of harm desiderata in \autoref{sec:literature}, to identify, typify, and mitigate harms stemming from new technologies. We next expanded the research to identify the types of harm relevant to forecasting within the established categories. This hybrid approach enabled us to explore the details of forecasting-related harms, eventually leading to an inductive thematic analysis. One benefit of using Microsoft Azure framework is its capability to facilitate discussions on harm across various levels, ranging from individual and community to organizational, societal, and environmental dimensions.

We used a semi-structured interview method; the detailed interview protocol can be found in \autoref{sec:interviewprotocol}. We used purposive sampling to ensure that we chose participants who could provide a wealth of information about the topic, prioritizing depth above mere numbers of interviews. We recruited people from academia and diverse organizations who have extensive expertise and experience in forecasting. These ``knowledgeable agents'' came from a variety of disciplines, including supply chain, business, public health, healthcare operations, humanitarian and development operations, government bodies, and software development (see \autoref{sec:sample}). This diverse representation enabled a thorough investigation of forecasting interests across organizational types and sectors.

We conducted 21 interviews, each lasting 40 to 65 minutes, all of which were completed online for convenience and accessibility. This produced about 1121 minutes of interviews, resulting in a rich dataset for analysis. Due to space constraints, only a sample quote from the interviews is provided in the \autoref{sec:appc}. However, the full interview transcripts will be made accessible via a GitHub repository.

Ethical approval for this study was obtained from the Cardiff Business School Research Ethics Committee (Approval ID: 1727) on 19 June 2023. The study adhered to Cardiff University's ethical principles as outlined in the Policy on the Ethical Conduct of Research Involving Human Participants, Human Material, or Human Data. Informed consent was obtained from all participants, data were anonymized after data collection, and measures were implemented to ensure confidentiality and data security.

\subsection{AI-assisted inductive coding method}

\begin{figure}[!htbp]
	\centering
	\includegraphics[width=.9\linewidth]{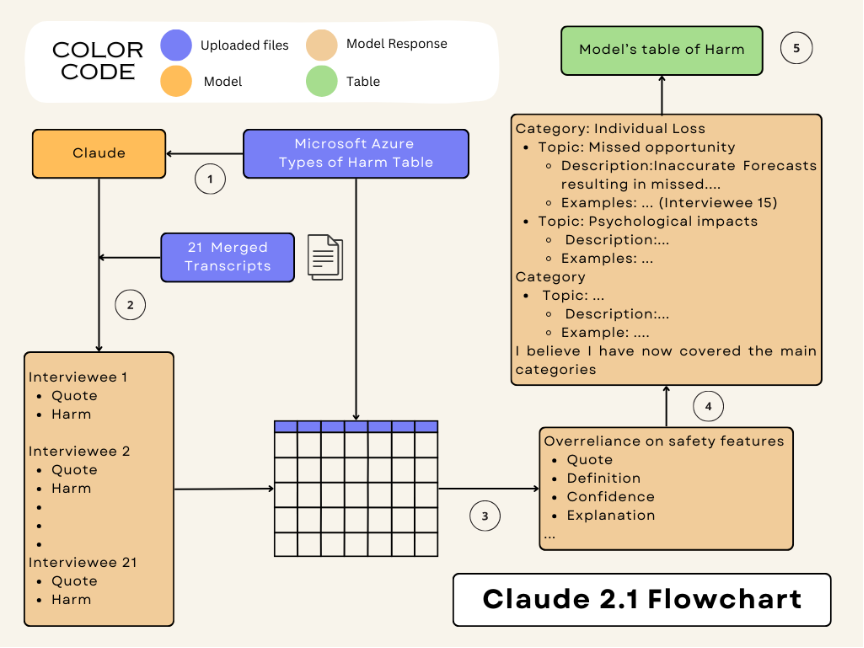}
	\caption{LLM use flowchart.}
	\label{fig:fig_llm}
\end{figure}

Inspired by the Microsoft Azure framework's taxonomy for responsible innovation, we combined a human-led inductive coding scheme with an AI-driven analysis centered on the extraction of key themes of harms in forecasting. The AI-driven analysis utilized a LLM (Claude 2.1) to help identify categories of harm from the 21 interviews' transcripts. Before supplying Claude the transcripts, we provided it with the Microsoft Azure framework as an example of a harm typology. We then requested Claude to categorize the harms mentioned in the interviews into the Microsoft Azure framework categories, providing relevant quotes from the interviews. Finally, we requested Claude to create a new framework of forecasting harms, along with relevant quotes. In all cases, we carefully inspected each suggested harm category, and validated each quote from the interview transcript provided by the LLM. See \autoref{fig:fig_llm} for a schematic of the process of the steps in the AI-driven analysis. \autoref{tab:my_label} shows the prompts used.

\begin{table}[!htbp]
	\caption{Prompts given to Claude LLM}
	\label{tab:my_label}
	\centering
	\begin{tabular}{p{4.5cm}p{12cm}}
		\toprule
		{\bf Action/Purpose}
		 & {\bf Prompt}
		\\
		\midrule
		Uploaded Microsoft Azure table (ManualTable.docx)
		 & This is how different types of harm are assessed under the Microsoft Azure framework
		\\
		\midrule
		Uploaded the document with 21 transcripts
		 & This is a merged transcript file of 21 interviews. Each interview talks about the potential harms involved in forecasting. Without using the table, what examples do you consider harmful in this transcript for each interviewee? Please show me the quote and the definition of the harm you identified for all interviewees.
		\\
		\midrule
		Match model response to Microsoft Azure table
		 & If you were to categorize this into the table I just uploaded, which “Topic” column would you put your response in? How confident are you towards each classification? Which of the harms do you identify that does not fit into the table? I would like you to include the definition and quote in each of your explanations as well.
		\\
		\midrule
		Create forecast harm framework using the model
		 & If we were to create a new framework for forecasting harms like ManualTable.docx, what are the themes, descriptions, and definition of the types of harms you would consider from all the interviewees' responses. I would also like you to specify the examples across all the interviewees that makes the type of harm presented relevant to this new framework. You do not need to use ManualTable.docx but it is provided to give you some ideas about the concept of harms. Remember to consider ALL the interviewees given in the transcript.
		\\
		\bottomrule
	\end{tabular}
\end{table}

\subsection{Human-led thematic analysis}

In parallel, and independent from the AI-driven analysis of the interviews, the first author conducted a thematic analysis based on the topics and harm categories outlined in the Microsoft Azure framework and led the coding of the interview data, beginning with an analysis of the interview transcripts using NVivo 12, which is a commonly used software package for analyzing semi-structured interview data. Relevant quotes were assigned to each harm category, with existing themes being confirmed and potential new categories explored. The findings from the AI-driven analysis were then compared and consolidated with those from the human-led thematic analysis. This synthesis was used to inform both the findings and the discussion. For categories identified by both analyses, relevant quotes were collected, while new themes identified by the human analysis were highlighted. A table summarizing these findings can be accessed in the supplementary materials.

\section{Findings: typology of harm in forecasting}
\label{sec:typology}

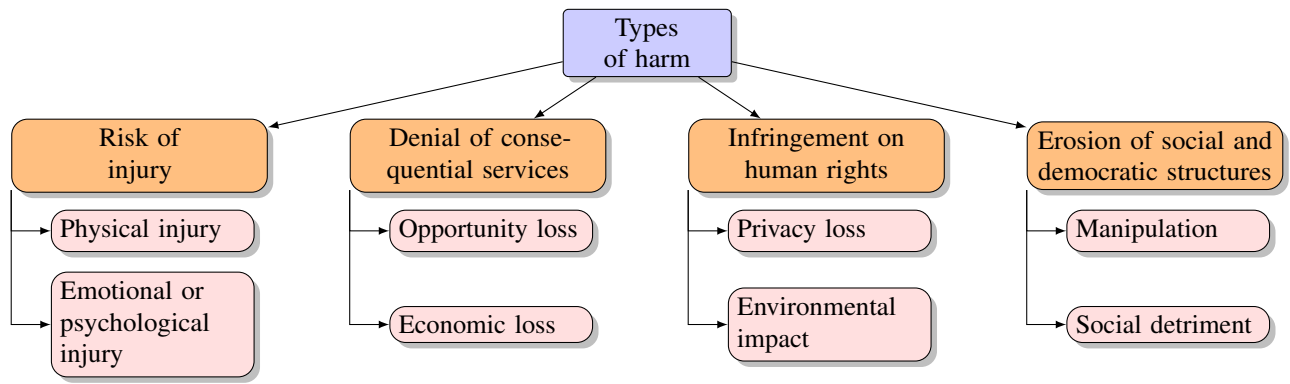
\begin{figure}[!ht]
	\centering\small
	\begin{tikzpicture}[
			level 1/.style={sibling distance=45mm},
			edge from parent/.style={->,draw},
			>=latex]
		\node[root] {Types of harm}
		child {node[level 2] (c1) {Risk of\\ injury}}
		child {node[level 2] (c2) {Denial of consequential services}}
		child {node[level 2] (c3) {Infringement on human rights}}
		child {node[level 2] (c4) {Erosion of social and democratic structures}};
		\begin{scope}[every node/.style={level 3}]
			\node [below of = c1, xshift=5pt] (c11) {Physical injury};
			\node [below of = c11, yshift=-7pt] (c12) {Emotional or psychological injury};
			\node [below of = c2, xshift=5pt] (c21) {Opportunity loss};
			\node [below of = c21, yshift=-7pt] (c22) {Economic loss};
			\node [below of = c3, xshift=5pt] (c31) {Privacy loss};
			\node [below of = c31, yshift=-7pt] (c32) {Environmental impact};
			\node [below of = c4, xshift=5pt] (c41) {Manipulation};
			\node [below of = c41, yshift=-7pt] (c42) {Social detriment};
		\end{scope}
		\foreach \value in {1,2}
		\draw[->] (c1.195) |- (c1\value.west);
		\foreach \value in {1,2}
		\draw[->] (c2.195) |- (c2\value.west);
		\foreach \value in {1,2}
		\draw[->] (c3.195) |- (c3\value.west);
		\foreach \value in {1,2}
		\draw[->] (c4.195) |- (c4\value.west);
	\end{tikzpicture}
	\caption{Types of harm in forecasting based on our interviews.}
	\label{fig:harmtype}
\end{figure}

This section captures the types of harm identified in the interviews. The analysis identified four main types of harm as illustrated in \autoref{fig:harmtype}: risk of injury, denial of consequential services, infringement on human rights, and erosion of social and democratic structures. In the following sections, we will highlight quotes from the interviews that illustrate each type of harm within these categories, stemming from inaccurate forecasts, poorly communicated forecasts, or simply publishing forecasts.

\subsection{Risk of injury}
This section explores how forecasting can potentially harm people, create hazardous environments, or contribute to significant emotional and psychological distress.
\subsubsection{Physical injury}
\begin{description}
	\item[Inadequate fail-safes] Forecasts of natural disasters can have dire consequences. In the case of hurricane forecasts: ``There have been instances in the last few years where they just simply got the number wrong. The hurricane took a turn they weren’t projecting'' (P2). Similarly, earthquake forecasting has proven challenging: ``Earthquakes are almost impossible to forecast, but if a forecast says we’re safe and an earthquake occurs, it can cost thousands of lives'' (P14). In some cases, the issue is not the failure to forecast the event, but a failure to forecast its impact, and plan accordingly: ``You produce a forecast of a potential fire without any plans in place to mitigate the effects, and that can effectively amplify the damage'' (P19). Similarly, during the COVID-19 pandemic, poor forecasting led to delayed governmental responses: ``If the government made a decision based on a poor forecast, you could measure the extra hospitalizations in hindsight'' (P21).

	\item[Exposure to unhealthy agents] When forecasts fail to forecast public health crises or environmental hazards, people may be left unprepared for the consequences, leading to substantial physical harm. One healthcare expert described the potential harm posed by inaccurate forecasts in family planning: ``Women who rely on short-term contraceptive methods face unplanned pregnancies if forecasts fail to deliver their needed protection on time'' (P8). Another participant recalled the L'Aquila incident: ``In Italy, some forecasters were jailed for failing to forecast an earthquake that led to loss of life'' (P6).
\end{description}

\subsubsection{Emotional or psychological injury}

\begin{description}
	\item[Over-reliance on automation and ML] Automated forecasting tools can foster blind trust in automated systems, leading to errors that are difficult to detect and correct, particularly when uncertainty is ignored or human judgment is replaced entirely. One major issue is that ``people tend to only look at the deterministic forecast\footnote{A point forecast that is represented as a single value summarizing a distribution. Alternatives are an interval estimate that gives a range of likely values, or a full probability distribution.} and discard the uncertainty bounds that are around it'', (P1) which can result in decision-makers overlooking the potential range of outcomes and the associated risks. Furthermore, automation can foster overconfidence, especially with point forecasts that present a single figure, giving the impression that ``there's a lot less uncertainty or risk associated with a forecast than there really is'' (P6). Another concern is that full automation driven by machine learning could eliminate the need for human forecasters altogether, which alarms some stakeholders. The idea of ``an enterprise that's completely driven by cognitive automation and then we don't need forecasters anymore'' (P17) raises fears of losing critical human insight, potentially leading to harmful decisions without proper oversight.  

	\item[Distortion of reality or gaslighting] Forecasts can mislead stakeholders, causing them to act on inaccurate information, leading to mistrust and harmful decisions. One participant noted, ``If you are forecasting processes with public effects, like healthcare or migration, the harms could range from panic to damaging public reactions'' (P14).

	\item[Reduced self-esteem/reputation damage] Reputational damage can result from both incorrect and correct forecasts. One participant shared: ``If you put a forecast out there and it’s wrong, you can suffer a reputational risk'' (P5). Even correct forecasts may cause harm if misinterpreted: ``You can lose credibility not through poor science, but through an inability to effectively communicate it to decision-makers'' (P21). This highlights the importance of both accuracy and communication in maintaining trust in forecasting.

	\item[Addiction/attention hijacking] This harm may be caused by excessive dependence on frequent short-term forecasts. One expert noted that planners often end up ``firefighting'', constantly reacting to the latest forecast without stepping back for long-term planning, which leads to stress and reactive decision-making (P11).

	\item[Anxiety, stress, and emotional distress] Forecasting uncertain or negative outcomes can cause significant emotional distress. A notable example is the COVID-19 pandemic, where inaccurate forecasts had widespread mental health impacts. As one participant explained: ``We went through all of the economic harm, mental health harm, and the impact on medical services due to incorrect forecasts about lifting lockdowns too early'' (P4). Similarly, healthcare forecasts can have a profound emotional impact: ``An accurate forecast of an unavoidable illness can ruin your last few years of life'' (P6). Another participant illustrated the potential emotional harm in everyday situations: ``When I go shopping in the evening and I can’t find strawberries anymore, there is emotional distress'' (P2). Thus forecasting, even in less critical domains, can evoke anxiety and frustration when expectations are not met.
\end{description}

\subsection{Denial of consequential services}
This section discusses how forecasting can limit access to essential resources, services, and opportunities, or result in economic losses.
\subsubsection{Opportunity loss}
\begin{description}
	\item[Insurance and benefit discrimination] Inaccurate forecasts can lead to unintended consequences in identifying and assisting vulnerable populations: ``Mapping poverty or any type of aid allocation can cause harm because imperfect models fail to perfectly identify the most vulnerable people, resulting in inefficient use of aid'' (P3). Another participant pointed out that this harm could extend to denying access to life-saving products: ``If a forecast shows there isn’t sufficient market, it may disallow access to a product that could improve health or save lives'' (P9).

	\item[Digital divide/technological discrimination] Marginalized communities can be further disadvantaged by being excluded from forecasting data. One interviewee stressed, ``Marginalized communities might be excluded from the data because they’re harder to reach, leading to decisions that overlook these populations'' (P21). This can exacerbate existing inequalities, leaving certain groups without access to critical services or benefits.

	\item[Loss of potential investment] Inaccurate forecasts may lead businesses to overestimate or underestimate market demand. One expert explained: ``If a manufacturer overproduces based on inaccurate forecasts, they risk losing confidence in the market and may not invest in promising public health products in the future'' (P9). In other cases, overly complex forecasting models that yield little business improvement may waste valuable resources: ``The harm lies in the opportunity costs of resources invested in a forecasting method that didn’t significantly improve outcomes'' (P2).
\end{description}

\subsubsection{Economic loss}
\begin{description}
	\item[Economic exploitation] Detrimental decisions may be based on flawed forecasts, such as over-committing resources or accepting poor employment conditions:  ``For long-term investments, inaccuracies in forecasts can hurt the business, especially when big investments are made in infrastructure like data centers'' (P18).


	\item[Misdirection of resources] Inaccurate forecasts also lead to wasted time, money, and effort. This issue was particularly evident in the construction and supply chain industries, where inaccurate forecasts resulted in the misallocation of materials and resources: ``The easiest harm to identify is the misdirection of resources due to forecast inaccuracies, especially when cognitive biases like optimism bias affect judgmental forecasting'' (P6). This misdirection extends to scenarios where pre-positioned resources for emergency response end up being in the wrong place, causing delays and increased costs (P5).

	\item[Financial loss] Several experts emphasized the financial damage caused by overproduction or underproduction, leading to excess inventory or lost sales. ``If forecasts are too optimistic, companies may overbuy inventory, leading to markdowns or waste'' (P17). This financial burden extends across industries, from manufacturing to tech, with one participant describing a situation where billions of dollars in unused servers were wasted due to inaccurate demand forecasting (P18). The economic harm also extended to job losses and reduced competitiveness: ``If a company can’t sell what it produced, it harms their competitiveness and could lead to layoffs'' (P11).
\end{description}

\subsection{Infringement on human rights}

Flawed forecasting models can lead to breaches of personal freedoms, mismanagement of private data, and environmental degradation.

\subsubsection{Privacy loss}
\begin{description}
	\item[Interference with private life]
	      As more time series data is generated by vehicles, as well as individual humans equipped with wearable sensors, new opportunities for harm arise. Forecasting models that misuse personal data can infringe upon privacy rights by revealing sensitive information.  While these mobility data are usually collected with benevolent intent, they can be repurposed in harmful ways: ``The data, especially because it’s at the individual level, is arguably being used for other purposes. There are even documented cases of people being persecuted or abducted because of this data'' (P3). The ability of forecasting models to misuse personal data underscores the importance of strict data privacy regulations, particularly when individual lives and rights are at stake.

	\item[Loss of freedom of movement or assembly]
	      Publishing forecasts of migration trends and political protests can cause governments to take preemptive measures to suppress them. In this way, forecasts risk ``causing the thing you're predicting to happen'' (P5), demonstrating the challenge of balancing the need for forecasting with their potential risks to freedoms.
\end{description}

\subsubsection{Environmental impact}
\begin{description}
	\item[Exploitation or depletion of resources]
	      Forecasting is frequently employed in the energy sector, where over/under-forecasts of energy demand can cause serious, but often, asymmetric harms at the societal level. Energy forecasts can also have indirect impact on the environment by shifting investments in other types of energy such as renewable, nuclear, etc. More directly, forecasting can also lead to significant environmental impact, particularly through the exploitation or depletion of resources. Over-forecasting, for example, can lead to the over-extraction of resources, with significant environmental and economic consequences: ``If you can get more accurate forecasts, there will be less waste, less things going to landfill'' (P6). Another participant noted that when products are over-forecast and over-procured, ``There is then the whole issue of waste disposal. How do you destroy products? Not just the products themselves, but also the packaging and everything that goes along with that'' (P9).

	\item[Waste] Inaccurate forecasts can lead to waste, particularly when excess products are procured that cannot be sold: ``When you get more than you sell, you need to go through markdowns, and for some items, they will just become waste because they are perishable'' (P10). This problem is widespread, affecting industries ranging from pharmaceuticals to retail: ``We’ve seen national governments over-forecast and over-procure products, which leads to large quantities of expired or unused products needing to be destroyed'' (P13).
\end{description}

\subsection{Erosion of social and democratic structures}

This section examines how inaccurate or biased forecasting can erode social and democratic structures, focusing on two types of harm: manipulation and social detriment. The findings indicate how forecasting can be used to mislead or exploit people, reinforce stereotypes, and potentially marginalize underrepresented communities.

\subsubsection{Manipulation}

\begin{description}
	\item[Misinformation]
	      Forecasts can spread misinformation by presenting biased or incomplete information, often influencing public perception and decision-making: ``forecasts [can be] used for political reasons to sway decisions, sway policy, or sway people’s opinions” (P12). This intentional use of misleading forecasts can cause significant harm, if people use these forecasts without understanding the underlying uncertainties or biases. Misinformation can also foster panic or misguided actions: ``They said, `The scientists have told us it’s going to be terrible,’ with no nuance or layering of that. And it had the risk, I think, in the media of causing some panic'' (P21).

	      Forecasts can also create a self-fulfilling effect, where public belief in a forecast influences outcomes: ``I think one potential risk is that people might accuse you of having caused the thing to happen, or that there’s this danger of making things a self-fulfilling prophecy'' (P5). This highlights how forecasts can shape societal behavior in ways that inadvertently validate or invalidate their forecasts.

	\item[Behavioral exploitation]
	      Forecasts based on behavioral patterns may exploit personal habits or biases, guiding people to make decisions that align with the forecast rather than their best interests. In some cases, forecasts are intentionally leveraged to drive particular behaviors. One participant explained, ``There are people who use forecasts or media coverage of high-profile incidents to sell the message, such as vaccinations, regardless of how ill-founded or incorrect the forecasts are'' (P4).

	      Furthermore, self-destructive forecasts can also harm democratic processes, such as voting behavior: ``A classic example is opinion polls where you forecast that one party is going to win an election, and voters decide not to bother voting\dots\ Opinion poll forecasts have been banned in some countries because of their danger of influencing the electorate, and possibly damaging democracy'' (P6).
\end{description}

\subsubsection{Social detriment}
\begin{description}
	\item[Stereotype reinforcement and loss of representation]
	      Forecasts that rely on historical data can perpetuate harmful stereotypes, particularly in fields like employment, education, and criminal justice. Such forecasts may reinforce existing biases and contribute to systemic inequality: ``If our data is on only a subset of the population, and then we forecast that \dots\ marginalized communities might be excluded from the data because they’re harder to reach'' (P21). Similarly, forecasts can lead to the loss of representation for minority or marginalized groups. Forecasting models based on aggregated data may obscure the specific needs of these groups, leading to decisions that fail to account for their unique circumstances: ``If our data is only on a subset of the population \dots\ it can set up socioeconomic or marginalized communities to be excluded'' (P21). Thus, overly simple forecasting methods can marginalize vulnerable populations, rendering their needs invisible in decision-making processes.
\end{description}

\section{Discussion}
\label{sec:discussion}



\subsection{Potential harm from forecasting}

\begin{figure}[h!]
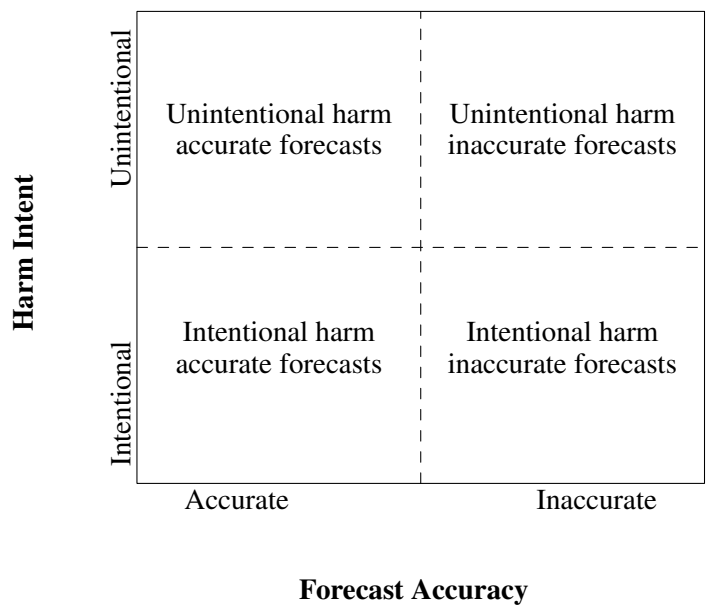

	\centering\vspace*{-1.5cm}
	\input harm_matrix.tex
	\caption{Forecast harm matrix}
	\label{fig:harmmatrix}
\end{figure}

Our analysis of the interviews suggests four broad types of forecasting harm, depending on the intention of the forecasters and the accuracy of their forecasts, as shown in \autoref{fig:harmmatrix}. Each has different ethical implications, highlighting the complex ethical and practical challenges in forecasting. For instance, one conflict is the probabilistic nature of forecasting harms and ethical practices that ascribe moral blame based on an actor's intent to cause harm. 

\begin{description}
	\item[Unintentional harm from inaccurate forecasts]
	      Unintentional harm can arise when forecasters produce poor forecasts, whether through a lack of suitable data, inadequate training, the intrinsic noise in the process, or a combination of these factors. For example, a forecaster provides a forecast for a humanitarian crisis, but it does not accurately forecast the hardest-hit areas. As a result, aid is misallocated, and vulnerable populations do not receive the necessary resources. One interviewee explained: ``your models are going to be imperfect \dots\ you're creating some indirect harm because you didn't perfectly identify the most vulnerable people.''

	      Further analysis might seek to identify criteria for evaluating the causes of forecast inaccuracy. If the inaccuracy stems from oversight, or negligence in the data collection or model-building process, forecasters or their organizations might be held legally or morally responsible for resulting harms. But if forecasters took all reasonable precautions in developing their forecasts, they might be released from liability for any harms that actually result. This roughly  captures what eventually happened to the seismologists in the L'Alquila case, and more generally parallels how determinations of malpractice are made in medical contexts. These considerations illustrate the need for more work aimed at integrating  forecasting-specific practices into traditional theories of harm in law and ethics.

	\item[Unintentional harm from accurate forecasts]


	      Accurate forecasts, while informative, can unintentionally inflict harm due to their impact on human behavior and emotional well-being. For instance, a healthcare forecast predicting the progression of an inevitable illness or genetic condition may cause severe psychological distress, creating ethical dilemmas for individuals and their families and often diminishing quality of life. Similarly, an accurate recession forecast can prompt consumers and businesses to cut spending preemptively, inadvertently accelerating the economic downturn it aimed to address. In both cases, the accuracy of these forecasts does not mitigate their potential to cause harm through unintended consequences on behavior, underscoring the complex role of forecasting in high-stakes domains. This scenario suggests that even with good intentions, forecasters in sensitive fields must carefully consider the broader impact of their predictions, particularly in areas where public trust and well-being are at stake.

	\item[Intended harm from inaccurate forecasts]
	      An actor or organization can knowingly publish an inaccurate forecast, which then indirectly inflicts damage due to its likely effects on human behavior --- akin to yelling ``fire'' in a crowded movie theater. For example, a forecaster may knowingly publish a false forecast suggesting that COVID-19 cases are decreasing and that the virus is no longer a threat, despite clear evidence to the contrary. As a result, people stop wearing masks and taking precautions, leading to a new wave of infections and deaths. Alternatively, a stock price forecaster may deliberately forecast a high price in order to inflate the price before selling their stocks. Fortunately, known cases of intentional forecasting harm are rare. Yet one interview participant explained that forecasters can knowingly manipulate the data or the forecast, creating an intentional disconnect between the forecast and the actual expected outcomes, with the conscious goal of inflicting harm.

	      This scenario could be viewed as motivation for professionalization and licensing of forecasters, at least for those working in high-stakes domains or who occupy institutional roles that rely on public trust and support. Furthermore, the intentional harm scenario suggests drawing on a similar distinction made in law between civil and criminal wrongs. Our study has implicitly assumed most forecasting harms are unintentional. The possibility of bad actors inflicting intentional forecasting harms could motivate the need for criminal sanctions on behalf of the state as a means of protecting society's interests.

	\item[Intended harm from accurate forecasts]
	      Even accurate forecasts can lead to harm if they are misused by malicious actors. While a forecaster may have good intentions, their forecasts can be exploited by adversaries for harmful purposes. As one interviewee explained, ``you could imagine that other war parties\ldots use this in a different way.'' For example, a forecast population displacement due to conflict is intended to guide humanitarian efforts to provide aid. However, hostile forces use the same forecast to target vulnerable populations, attacking areas where displaced families are expected to gather. Another example could occur in military scenarios, when accurate forecasting models are used to plan and time strategic disruptions to an enemy's supply chain \citep{layton2021fighting}.

	      However, the majority of risks arise from the complexities of forecasting, rather than deliberate malice. Interviewees emphasized that forecasters generally aim to produce the most accurate forecast possible. Yet inherent limitations of the forecasting process, imperfect data, or unforeseen events, can still lead to unintended negative consequences.
\end{description}

\subsection{Domains prone to harm}

AI researchers and legal scholars find it useful to designate applications of AI as falling into discrete risk tiers. More generally, the legal concept of \emph{proportionality} relates a risk tier to a standard of safety or scrutiny \citep{karliuk2023proportionality}. Higher risk applications require proportionally higher levels of scrutiny. We see a similar need to determine risk-tiers for domains where forecasting is commonly applied. Forecasts may cause harm in sensitive domains where human lives, financial stability, and societal well-being are at risk. Healthcare, humanitarian and environmental crises, economics, politics, and other domains where vulnerable populations reside are especially prone to harm, as forecasts can influence critical decisions that have lasting effects on many people's lives.

One interviewee pointed out how politics and healthcare came immediately to mind as domains where incorrect forecasts can cause substantial harm, because these areas often affect large populations and drive policy decisions. Politics directly shapes societal responses, influencing everything from public confidence to policy interventions. So election forecasts, or forecasts about policy decisions, have high harm potential. In healthcare, forecasting errors can have life-and-death consequences. For example, if a pandemic forecast underestimates the spread of disease, healthcare systems may be unprepared, leading to preventable deaths and long-term public health crises. Misleading or overly optimistic forecasts in these sectors can lead to inadequate preparation and insufficient responses, which can ripple through societies.

Humanitarian work is also a domain where the stakes are exceptionally high. Forecasts in humanitarian contexts, whether they involve natural disasters, migration flows, wars, or droughts, are critical for organizing relief efforts and protecting vulnerable populations. As one interviewee noted, in humanitarian contexts, ``the stakes are just higher'' because the margin for error is slim. A flawed forecast can result in misallocated resources, delayed interventions, or missed opportunities to provide life-saving assistance, potentially with devastating consequences for those in need. 

Forecasts that involve vulnerable populations---whether in humanitarian crises, economic downturns, or migration flows---are fraught with risk. Vulnerable populations often rely on external support, and inaccurate forecasts can exacerbate an already precarious situation. If a forecast underestimates the needs of a vulnerable population, the resulting lack of support can deepen the crisis. One interviewee remarked that when dealing with vulnerable populations, the opportunity for harm is even greater, as they have fewer resources to adapt or recover from forecast-related decisions. This is especially true in contexts where the forecast shapes policy or public intervention, such as migration or healthcare. The discussion of vulnerable populations highlights the need for greater clarity around the social role and responsibilities of forecasters. In biomedical ethics \citep{beauchamp2001principles}, special considerations apply when experimental participants belong to vulnerable populations, such as the homeless, political refugees, or those suffering from mental illness. And in law and medicine, exploitation of the vulnerable is to some extent mitigated by professionalization, codes of ethics, and community and evidence-based quality of care standards. 

The potential for harm also extends to economic and financial forecasts, which can cause widespread disruption and loss. Financial forecasts shape decisions in markets and firms, influencing everything from inventory orders to asset pricing. As one participant pointed out, an incorrect financial forecast can lead to significant economic loss, whether through overpricing or underpricing assets, or causing market disruptions that impact the broader economy. Additionally, economic forecasts that shape public policy, such as those predicting inflation or employment trends, carry risks for the entire population, particularly if they inform policy decisions that fail to address underlying economic vulnerabilities.

An important thread running through these domains is that harm from forecasting is not only linked to the forecast itself, but also to its public perception and the societal response. In areas like healthcare, migration, and even economic forecasting, the way people react to forecasts can amplify or mitigate the harms. As one interviewee explained, ``harms are associated with the sort of response that the public tends to have to such forecasts.'' This underscores the behavioral and psychological consequences of a forecast that go beyond its predictive accuracy. For example, a public panic in response to a pessimistic migration forecast can lead to hasty political decisions, potentially exacerbating the situation. Currently, forecasters receive little to no formal training or education about how forecasts may be psychologically perceived by an audience.

Ultimately, the societal impact of a forecast is closely tied to the domain it addresses. Forecasts related to elections, inflation, pandemics, or migration have far-reaching effects on policy-making and public sentiment. As one interviewee summarized, ``the potential harm of a forecast is directly related to the impact that forecast will have on human life.'' In these cases, even a small forecasting error can have serious consequences, from undermining public health efforts to causing financial instability or social unrest.

\subsection{Possible mitigation strategies to minimize harm in forecasting}

\label{sec:harmmitigation}

Our interviewees emphasized that minimizing harm in forecasting is a shared responsibility between forecasters and decision-makers. Furthermore, respondents suggested that to improve forecasting transparency and accountability, forecasters should clearly communicate model assumptions, uncertainties, and potential impacts, while taking ownership of their forecasts, particularly in high-risk domains. Moreover, several interviewees stressed the importance of understanding the audience and purpose of forecasts, raising concerns about misuse, and the value of using simulation-based approaches to illustrate potential outcomes. 

Conversely, the interviewees largely agreed that decision-makers should critically engage with forecasts, assess their limitations, and ensure they are fit for purpose. Both parties should collaborate to establish feedback loops, refine forecasts, and promote non-harmful use, while recognizing that forecasts are not absolute truths but epistemic tools that guide decisions within a broader context of uncertainty.

\begin{figure}[!htb]
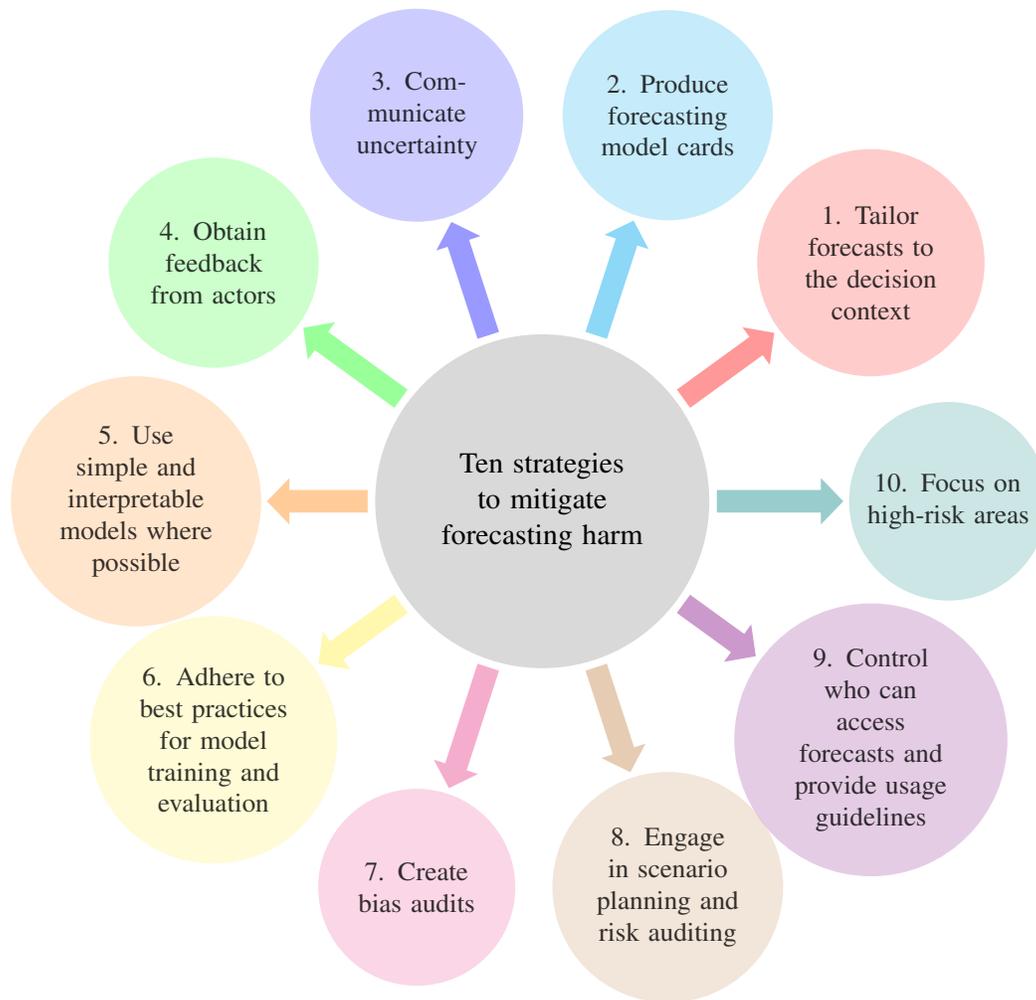

	\smartdiagramset{
	planet font=\normalsize,planet text width=28mm,
	satellite font=\small, satellite text width=22mm,
	distance planet-satellite=54mm,
	/tikz/connection planet satellite/.append style={-{Triangle[length=3mm,width=6mm]}, line width=3mm},
	}
	\centering
	\smartdiagram[constellation diagram]%
	{
		Ten strategies to mitigate forecasting harm,
		1. Tailor forecasts to the decision context,
		2. Produce forecasting model cards,
		3. Communicate uncertainty,
		4. Obtain feedback from actors,
		5. Use simple and interpretable models where possible,
		6. Adhere to best practices for model training and evaluation,
		7. Create bias audits,
		8. Engage in scenario planning and risk auditing,
		9. Control who can access forecasts and provide usage guidelines,
		10. Focus on high-risk areas%
	}
	\caption{Mitigation strategies extracted from interview analysis}
	\label{fig:mitigation}
\end{figure}

Drawing on these and other points, we have identified ten approaches to mitigate the harm that can arise from forecasting, which are illustrated in \autoref{fig:mitigation}. These can help forecasters reduce the likelihood of harm while ensuring forecasts are used appropriately by decision-makers. 

\begin{description}
	\item[1. Tailor forecasts to the decision context]
	      Interviewees emphasized the importance of tailoring forecasts to the specific needs and contexts of decision-makers. Forecasts that are not aligned with decision-making processes can cause harm, as stakeholders may misunderstand their intent or limitations. One participant explained, “we should communicate that these forecasts are made under certain assumptions, and the assumptions may be, for instance, that things will continue as they are now.” Another example highlighted the need for qualitative information alongside the quantitative, explaining the use case and time horizon so that decision-makers can properly apply the forecast. Close collaboration with decision-makers is essential to ensure that forecasts are tailored appropriately and that the assumptions and limitations are fully understood.

	\item[2. Produce forecasting model cards]
	      Forecasters should ensure that forecast consumers---the stakeholders--- know what the forecast represents to prevent potential manipulation or misinterpretation. Forecast stakeholders should have access to information about the data, models, assumptions, the conditions under which the forecast applies, and forecast variability. Such transparency enables stakeholders to properly interpret and trust the forecasts.  \emph{Model cards} \citep{mitchell2019model} could a viable solution to promote more standardized forecast reporting. Standardized reporting is used in clinical trials to summarize drug efficacy data and convey the uncertainty attached to regulatory decisions \citep{fischhoff2014communicating}. Forecast model cards could provide a standardized disclosure mechanism, not unlike food nutrition labels or  publicly available financial reports for firm investors, that help forecast stakeholders grasp the limitations and potential biases of the forecasts produced by a forecasting model. For instance, model cards could help guard against the file-drawer problem in forecasting, and mitigate the ``problem of many hands'' by encouraging accountability for forecasts by giving the name of an accountable individual or organization to be contacted in case of concern. Further, publicly available forecasting model information would allow stakeholders to engage in more informed decisions about where forecasting should and should not be applied, given the level of forecasting uncertainty and the costs of error.


	\item[3. Communicate uncertainty]
	      A consistent theme in the interviews is the need to communicate uncertainty clearly. One participant highlighted that “providing just one number is never good enough in decision-making systems.” Forecasters must express the uncertainty inherent in their models and provide a range of potential outcomes rather than a single point forecast. Without this, decision-makers may make incorrect assumptions about the certainty of the forecast, leading to misinformed decisions. Another interviewee emphasized, “Always speak to the assumptions you are making when you are doing the forecast and always talk about the uncertainty you have on your results,” noting that if uncertainty isn’t communicated, stakeholders often make their own, sometimes harmful, assumptions. Communicating uncertainty is critical to minimizing harm, particularly in fields where decision-making is sensitive to small changes in the forecast.

	\item[4. Obtain feedback from actors]
	      Several interviewees mentioned the need for feedback from decision makers. As one participant explained, “If you simply send your forecast off to somebody and never hear anything else from them, then you don’t know whether it’s caused harm or not.” Regular communication and feedback loops between forecasters and stakeholders are critical to ensuring that forecasts are used correctly and understood within the broader decision-making context. Consulting with a wide range of stakeholders, including those directly impacted by the forecast, helps forecasters identify potential issues early and adjust forecasts to meet user needs more effectively.

	\item[5. Use simple and interpretable models where possible]
	      Overly complex models can pose a risk if they are difficult for decision-makers to understand and apply. As one interviewee noted, “If the model is too complex and people are not using it, you’re also somehow doing harm.” Simpler models that are easier to understand are sometimes more effective than sophisticated ones, particularly when decision-makers are unfamiliar with complex statistical techniques. Ensuring that models are interpretable and accessible increases the likelihood of adoption and appropriate use, avoiding harms that arise through lack of trust in the forecasts.

	\item[6. Adhere to best practices for model training and evaluation]
	      Adhering to best practices in model training and evaluation is critical to ensuring forecast accuracy and minimizing harm. One interviewee highlighted that accuracy is closely tied to reducing harm: “If accuracy is related to reduction of harm, then a recommendation would be to follow best practices in both the training of the models and the evaluation of the models.” Additionally, forecasts should be benchmarked against simpler approaches to ensure they provide meaningful improvements. Proper evaluation helps to avoid reliance on overly complex models that may not offer substantial benefits over simpler alternatives.

	\item[7. Create bias audits]
	      Bias in forecasts can lead to harm, particularly when certain populations or perspectives are neglected. Interviewees noted the importance of auditing models for biases at each stage of the forecasting process. Furthermore, diverse teams can help mitigate blind spots and identify potential harms that a homogeneous team might overlook. One interviewee emphasized that “having a diversity of perspectives” on the forecasting team is crucial to detecting problems early and ensuring that forecasts serve all affected groups fairly.

	\item[8. Engage in scenario planning and risk auditing]
	      Engaging in scenario planning and risk audits can help forecasters anticipate potential harms. One interviewee suggested that, especially in conflict-affected areas, forecasters should “think really out of the box” and engage in exercises to assess what wrong could happen if a forecast is published. By proactively identifying risks, forecasters can implement safeguards to reduce the likelihood of harmful outcomes.

	\item[9. Control who can access forecasts and provide usage guidelines]
	      Controlling who has access to forecasts and how they are used is another potential mitigation strategy. One interviewee suggested that published forecasts “shouldn’t just be consumed by anyone for any reason,” arguing that access control mechanisms can help ensure that forecasts are used appropriately. Proper onboarding processes and close monitoring of how forecasts are applied can help prevent harmful misuse. Granularity is also important---forecasters should provide clear guidelines on how to use forecasts at different levels of aggregation (e.g., weekly versus monthly) to ensure that users apply them correctly.

	\item[10. Focus on high-risk areas]
	      Another strategy emphasized was the need to prioritize areas where forecasting could cause the most harm. One interviewee used the analogy of “focusing on the tigers and not the mice,” meaning that forecasters should concentrate on areas where errors would have significant consequences, such as high-stakes sectors like healthcare, supply chains, and humanitarian responses. Not all forecasts carry the same level of risk, and focusing on areas with the greatest potential for harm allows forecasters to minimize the likelihood of serious negative outcomes. By identifying the most critical areas, forecasters can work to improve accuracy and relevance where it matters most.

\end{description}

\section{A Research agenda}
\label{sec:FutureDirections}

Based on our preliminary study and interviews with forecasters, we describe some promising areas of future research.
\begin{description}
	\item[Compensation for forecasting harms]
	      Context matters in determining what constitutes a harm. Just as in clinical experiments, consent to the risks of harm could in principle morally legitimize harms stemming from forecasting. To date, however, the lack of research attention paid to forecasting harms suggests this consent has been largely tacit and presumed. Consent is especially relevant for vulnerable populations which often must make important decisions under duress. We hope our research can stimulate more transparent discussion among forecast stakeholders about which risks related to forecasting ought to be explicitly consented to, and which not.

	      Drawing on an example taken from climate change and the ethics of corrective justice, we offer one suggestion for navigating the complex tradeoffs between the benefits and risks of forecasting. Just as some corporations voluntarily pledge to offset their greenhouse gas emissions \citep{hyams2013ethics} (e.g., by planting new trees), organizations doing large-scale forecasting that creates relatively trivial harms, such as representational harms \citep{suresh2021framework}, or minor allocative harms related to mis-forecasting the supply/demand of certain non-essential items, can donate or contribute to society in other ways. The idea here is that if it is practically impossible to separate out the good from the bad effects of forecasting in society, then as an alternative, those bad effects might be morally mitigated by compensatory actions. 


	\item[Developing ``fair'' forecasting metrics]
	      A major area of forecasting research concerns the development and analysis of accuracy metrics \citep{hyndman2006another, davydenko2016forecast}. Yet most work in fair machine learning focuses on binary classification contexts. For instance, one measure of discrimination compares true positive rates across groups, a fairness metric known as \emph{equal opportunity} \citep{hardt2016equality}. This leaves practitioners working on numerical prediction problems with little guidance on what constitutes ``fair performance'' of a forecasting model. We encourage researchers interested in the ethics of forecasting and questions of distributive justice to develop ``fairness'' metrics suitable for forecasting contexts. For example, mean error can indicate systematically over- or under-forecasted values, but one may wish to evaluate how such biases are distributed across society. There may be valid ethical justifications for distributing the harms associated with these errors across society (and their associated social and business burdens) to address historical inequities. At the same time, we should caution against narrowly treating fairness as a property of forecasting algorithms, independent of larger social contexts in which forecasting models are deployed.

	\item[Explainability, transparency, and contestability of forecasts]
	      \label{sec:explainable}
	      Potential harm is reduced if decision-makers are provided with understandable explanations for highly sophisticated forecasts. AI researchers are increasingly focused on explaining the output of predictive models to non-experts using a variety of explainable AI techniques \citep{arrieta2020explainable}. Advances in this area have come partly in response to new legislation and growing consumer demand for explainable predictions, especially in high-stakes contexts related to finance, medicine, and education. A key question is whether more inherently interpretable but potentially less accurate methods be used, rather than more data-driven blackbox models \citep{rudin2019stop}. To date, most of the research and public discussion around transparency has centered on applications of supervised machine learning, but has not addressed the area of time series forecasting methods.

	      Political and legal researchers have focused on the importance of being able to contest automated decisions, and for holding people and organizations accountable, both legally and ethically, for when predictions go wrong. We believe that similar lines of research will be important in the ethics of forecasting, thus motivating our proposal for developing ``forecasting model cards'' which list accountable persons and organizations.

	\item[Addressing forecasting inconsistency]
	      Forecasting models trained on the same data can lead to different forecasts because the optimization methods often rely on random number generation and, by chance, randomly selected initial parameter values can converge to one of many possible local optima in a complex loss landscape. Other sources of inconsistency are the of choice of series length, training/holdout partitioning, performance measures, and even the software used. The resulting inconsistency in forecasted values can reduce end-user trust in forecasts as they may view the forecasted model as unreliable.

	\item[Managerial override of forecasts]
	      Many interviewers noted that attempts to adjust a forecast away from its initial value usually leads to more harm than benefit. One reason is managers not understanding the inherent noise in the process, and reacting to random variation rather than signal. Another reason is that there is often a lag between a corrective action and the observable consequences of that action. If, for example, a product is forecasted as having low demand, a marketing team may spend months developing an appropriate strategy to counter-act this forecast. But the marketing intervention will take time before its effects on demand are noticed. In other words, there is a temporal miscalibration between corrective action and outcomes that can lead to exacerbation of harms in certain cases.  As suggested in \citet{shmueli2016practical}, to address this issue, organizations can collect data on when such corrections occur and later compare these corrected versions against the original version when the actual data arrive. A similar proposal has been made in criminal justice risk assessments when judges override algorithmic recommendations \citep{koepke2018danger}. To date, little research has tried to explore the conditions under which managerial corrections ultimately help or hurt forecasting process.

          \item[Statistical education and training for managers]
	      Lastly, many interviewers noted that managers---a major audience for forecasts---often do not grasp the differences between point estimates and distributions, leaving them confused or angry when forecasts ``turn out wrong''. This reaction demonstrates a need for greater education in probability and statistics for business students and managers.
\end{description}

\section{Conclusions}\label{sec:conclusions}

Considerable research attention has focused on the harms of AI/ML, yet our understanding of forecasting harms remains underdeveloped. The special nature of forecasting pipelines, data, and applications makes it difficult to apply existing theories of harm and AI/ML harm frameworks to the forecasting context. To remedy this gap, we combined an AI-driven analysis with an inductive, human-led thematic analysis to identify four emergent themes of harm related to forecasting. These themes were synthesized with philosophical and legal insights about the concept of harm, as well as specific findings from research on the harms of AI/ML, to identify and taxonomize the harms of forecasting. We defined harm as the unjust defeating of the interests of an individual or collective such as a society, organization, or ecological system, during the practice of forecasting. This definition helped us develop a taxonomy of harm focused on the intent and accuracy of a forecasting model. Besides contributing a novel organizational framework for understanding harms relevant to forecasting, we also outlined several harm mitigation strategies.

Our study revealed that many of the potential harms of forecasting lie not just in the act of forecasting itself, but in the publication of the forecasts, and in how forecasts are used to inform behaviors and decisions that impact broader social and ecological systems. Whether in inventory management, financial markets, or public policy, the risks are directly tied to the actions taken based on those forecasts. The question is not whether a forecast is always right, but whether the net benefits of forecasting outweigh the risks. This is arguably a policy or value question that must be decided by particular political communities. Our harm taxonomy can promote a more comprehensive public deliberation regarding the nature of the trade-offs involved.

One related question concerns responsibility for the harms caused by forecasting. 
Whether a poorly produced forecast actually results in harm can depend on luck. This complexity may spur discussion within the forecasting community around the adequacy of existing ethical guidelines or help inspire new ``standards of care'' uniquely adapted to forecasting. A greater appreciation and recognition of the harms of forecasting may even motivate steps towards the professional licensing of forecasters, particularly those working in high-risk domains or during emergency periods such as a global pandemic. Indeed, professionalization could represent a broader, systemic approach to harm reduction that, for instance, involves institutionalizing some of the harm mitigation strategies discussed in Section \ref{sec:harmmitigation}.

Our focus on forecasting harms may tempt one to conclude that forecasting is inherently dangerous. This conclusion should be avoided. We believe that while forecasting can lead to harm, abandoning it completely would not eliminate risk---it would only make us blind to it. If organizations were to completely abandon forecasting, this would likely lead to greater uncertainty and inefficiency in many important processes throughout society. Forecasting, even when imperfect, offers a mathematically sound basis for decision-making that is often superior to human intuition. For example, without forecasts, supermarkets might overstock products, leading to waste and inefficiency, as one interviewee explained. The key is not to stop forecasting altogether, but to focus on identifying and mitigating potential harms, with the goal of gradually moving towards more responsible forecasting practices. 

This work therefore aims to provide an empirical grounding for the development of more reflexive and responsible forms of forecasting in society. Reflexivity is an important aspect of science and responsible innovation \citep{stilgoe2013developing}. Responsible forecasting requires critically self-examining the activities, commitments and assumptions on the part of actors and institutions who use forecasting. Having a nuanced and open discussion of forecasting harms may be uncomfortable, especially when practical questions of legal liability arise. Still, the alternative of moral disengagement appears to us an even less attractive option. As one interviewee stated, ``forecasting is not just an [abstract] exercise\dots\ there are [concrete] implications.'' Taxonomizing and recognizing the specific harms stemming from forecasting can help organizations take initial steps towards becoming more responsible, transparent, and effective in minimizing and preventing the inherent risks of forecasting.

\section*{Declaration of Conflicting Interests}

The author(s) declared no potential conflicts of interest with respect to
the research, authorship, and/or publication of this article.





\singlespacing
\bibliography{literature}

\FloatBarrier\newpage
\appendix

\includepdf[pages=1, scale=.95,pagecommand=\section{Interview protocol}\label{sec:interviewprotocol}]{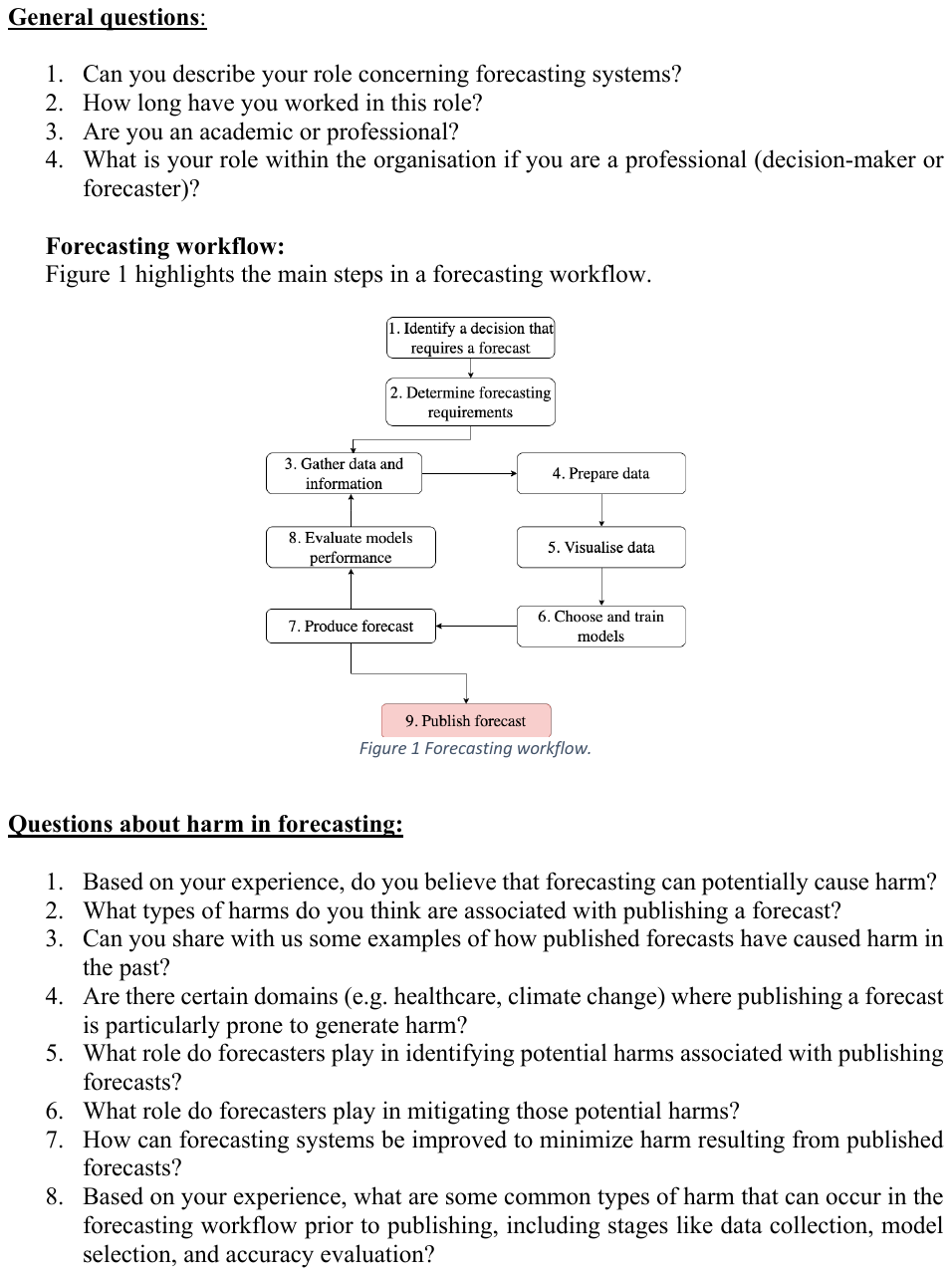}

\section{Summary of participants}\label{sec:sample}

\begin{table}[!htb]
	\caption{Summary of participants' domain, expertise and interview length}
	\centering\tabcolsep=0.1cm\small
	\begin{tabular}{llllc}
		\toprule
		\textbf{Code} & \textbf{Perspective} & \textbf{Role}                           & \textbf{Sector}                         & \textbf{Interview length} \\
		              &                      &                                         &                                         & (minutes)                 \\
		\midrule
		P1            & Practitioner         & Customer manager                   & Software vendor  & 61                        \\
		P2            & Practitioner         & Data Scientist                          & Retail                                  & 55                        \\
		P3            & Academic             & Professor                               & Computer Science                            & 55                        \\
		P4            & Academic         & Research Fellow            & Population and Public health                             & 53                        \\
		P5            & Practitioner        & Emerging Technologies lead                         & Goverment                           & 60                        \\
		P6            & Academic             & Professor                               & Business                                & 54                        \\
		P7            & Practitioner         & Marketing manager                                      & Software vendor                         & 64                        \\
		P8            & Practitioner         & Project Manager                         & Public health                          & 62                        \\
		P9            & Practitioner         & Senior analyst                         & public health supply chain                     & 47                        \\
		P10           & Practitioner         & Data Scientist                                & Retail                        & 57                        \\
		P11           & Practitioner         & CEO                                     & Supply chain software                   & 57                        \\  
		P12           & Practitioner         & Director            & Business                              & 44                        \\
		P13           & Practitioner             & Analyst                               & Global public health               & 58                        \\
		P14           & Practitioner         & CEO                & Software company                        & 40                        \\
		P15           & Practitioner         & Demand planner                           & Pharmaceutical                          & 65                        \\
		P16           & Practitioner         & Scientific lead         & Humanitarian                     & 54                        \\
		P17           & Practitioner         & Director for data science and service innovation                                     & Software                   & 57                        \\
		P18           & Practitioner         & Product manager                         & Technology                       & 56                        \\
		P19           & Academic         & Professor                   & Business                           & 50                        \\
		P20           & Practitioner         & Data Scientist                          & Retail                         & 55                        \\
		P21           & Academic         & Professor                 & Infectious disease                            & 43                        \\
		\bottomrule
	\end{tabular}%
\end{table}

\newpage

\section{Sample evidence from the interview quotes}\label{sec:appc}
\footnotesize

\subsubsection*{Risk of Injury / Physical injury / Inadequate fail-safes}
\vspace*{-0.4cm}
\begin{longtable}[l]{@{}ABC@{}}
	\toprule
	\textbf{Description in Miscrosoft} &
	\textbf{Description adapted to forecasting} &
	\textbf{Quotes from interviews} \\
	\midrule
	Real-world testing may insufficiently consider a diverse set of users and scenarios.  &
	Forecasting models may lack mechanisms to account for unexpected conditions, leading to inaccurate predictions in diverse, real-world scenarios. &
	(P2) "Hurricanes are a big causer of damage, and it's mostly in the Southeast of the United States, and the representation of the hurricane forecasts they wisely show it as a conic -- the path with -- and certainly a range around it and so on. But even with that, it's still imperfect and there have been instances in the last few years where they just simply got the number wrong. The hurricane took a turn they weren't projecting."
	\par
	(P14) "Earthquakes in particular I think are almost impossible to forecast, but if a forecast is produced saying, 'No, we're reasonably safe for the next few years,' and an earthquake occurs, it can cost thousands of human lives."
	\\&&
	(P16) "So the actions that were taken were not strong enough, were not adequate."
	\\&&
	(P19) "So you produce a forecast of potential fire without any plans in place to mitigate the effects and that can just effectively amplify that."
	\\&&
	(P21) "if a government, say in COVID made a decision to respond or not respond based on a forecast. And if you could establish that that objectively was a poor decision in hindsight, say, and measure the extra number of hospitalisations, yes, it could still happen."
	\\
	\bottomrule
\end{longtable}

\subsubsection*{Risk of Injury / Physical injury / Exposure to unhealthy agents}
\vspace*{-0.4cm}
\begin{longtable}[l]{@{}ABC@{}}
	\toprule
	\textbf{Description in Miscrosoft}  &
	\textbf{Description adapted to forecasting}  &
	\textbf{Quotes from interviews}  \\
	\midrule
	Manufacturing, as well as disposal of technology, can jeopardize the health and well-being of workers and nearby inhabitants.  &
	Poor forecasting, especially in health or environmental sectors, could expose stakeholders to unforeseen risks, such as public health hazards or environmental crises. &
	(P7) ``Earthquakes in particular I think are almost impossible to forecast, but if a forecast is produced there saying, `No, we’re reasonably safe for the next few years,' and an earthquake occurs, it can cost thousands of human lives. In fact, some Italian – I wrote this in a book, but it’s a few years ago. I’ve forgotten the details. Some Italian forecasters actually went to jail because they failed to forecast an earthquake."
	\par
	(P6) ``I guess my context and experience is only in healthcare, and it’s probably a good domain where the harm affects lives. And so it’s not just about, `Well, I won’t have my laptop on time or my phone on time.' It’s about women who rely on a short-term contraceptive method, like they need it every two, three months. Or even emergency contraceptives where it is needed tomorrow, and that can affect unplanned pregnancies should women not have that protection"
	\\&&
	(P8) ``I think there was that case in Italy, the legal case of, was it earthquake or volcano issue where some scientists got in trouble for not effectively communicating the risk of that natural disaster happening. And perhaps there was a lo- – there may have been loss of life, loss of property because of that. So there are potential physical, environmental types of dangers from forecasting, from inadequately communicated or just poorly inadequately created forecasts.
	\\
	\bottomrule
\end{longtable}

\newpage

\subsubsection*{Risk of Injury / Emotional or psychological injury / Overreliance on automation}
\vspace*{-0.4cm}
\begin{longtable}[l]{@{}ABC@{}}
	\toprule
	\textbf{Description in Miscrosoft}   &
	\textbf{Description adapted to forecasting} &
	\textbf{Quotes from interviews}  \\
	\midrule
	Misguided beliefs can lead users to trust the reliability of a digital agent over that of a human.   &
	Blind faith in automated forecasting tools without proper validation may lead to significant errors, causing stakeholders to trust flawed predictions over expert judgment. &
	(P1) ``People tend to only look at the deterministic forecast and discard the uncertainty bounds that are around it which could be quite large."
	\par
	(P6) ``I think the first is over-confidence. I think particularly with things like point forecasts where you present a single figure, it can create the impression that there's a lot less uncertainty or risk associated with a forecast than there really is."
	\par
	(P17) ``So one thing that really scares some of our clients or some people that we work with is the full automation that machine learning can cause. Like the idea that an enterprise that's completely driven by cognitive automation and then we don't need forecasters anymore."
	\\
	\bottomrule
\end{longtable}

\subsubsection*{Risk of Injury / Emotional or psychological injury / Distortion of reality or gaslighting}
\vspace*{-0.4cm}
\begin{longtable}[l]{@{}ABC@{}}
	\toprule
	\textbf{Description in Miscrosoft}    &
	\textbf{Description adapted to forecasting}   &
	\textbf{Quotes from interviews}   \\
	\midrule
	When intentionally misused, technology can undermine trust and distort someone's sense of reality.   &
	When misused or manipulated, forecasts can distort stakeholders' perception of future events, leading to mistrust or harmful decisions based on inaccurate predictions. &
	(P1) ``We have a lot of not firefighting, but pre-firefighting [laughs] people who are -- who tend to overreact to a time change in the forecast."
	\par
	(P3) ``So for example do people then send more law enforcement or more -- you know, somehow try to curb undocumented migration into certain parts of the US more. So it's maybe slightly less direct physical harm but it still has an element of physical harm just showing presence of people in areas where they're not supposed to be, or maybe they're not known to be also."
	\\&&
	(P4) ``Right. And so the first one, working with public health, and especially public health in near real-time, they’re very risk averse. It's a cultural thing, and obviously given some of the data they work with it’s entirely understandable. But from their perception, so I haven’t been allowed to publish my near real-time flu forecasts in real-time. We’ve been able to publish them in academic papers after a flu season, and retrospective analysis then showed and talked about where the forecasts did and didn’t perform well. But the public health staff, for better or worse, have a default mindset that making anything available publicly is opening themselves up to criticism and just negative impacts."
	\\&&
	(P8) ``But if the purpose of the forecast is not understood, so for example like a short-term versus a long-term, you can take what was supposed to be a short-term forecast because of the data that was used. So let's say three to five years, but they take it and then go off with a business plan that could be like five to ten years in the making to see that demand be generalised into actual sales. And so there's a misunderstanding of how that forecast should be used, and so if you're thinking in short-term, `Where's my ROI?' the donors may not see that. And so there's just, I guess, general frustration in that aspect."
	\\&&
	(P9) ``They then really felt somewhat betrayed by the donor community and by feeling like they'd done their part. So that's more of the like trust, and then wanting to engage with this public health community moving forward and then feeling -- I assume it had detrimental affects to their business as well."
	\\&&
	(P10) ``It's really hard to justify why you have larger or small number based on the forecast, because it's not just trend and last year."
	\\&&
	(P14) ``If you are forecasting processes that have a policy or a public effect, like healthcare or even where we mentioned before in the conversation, like migration flow, then obviously the harms are associated with the sort of response that the public tend to have to such forecast, which might be ranging from panic to reacting in a certain way of the process."
	\\&&
	(P21) ``And they spoke publicly about that. They didn’t show the forecast because they were all kept in secret boxes. But they said essentially, `The scientists have told us it’s going to be terrible' with no nuance or layering of that. And it had the risk, I think, in the media of causing some panic."
	\\&&
	(P1) ``Maybe they're interested in forecasting migration flows not to facilitate the safe and orderly migration but basically to build higher walls and optimise pushback of boats or whatever it might be. So those, of course, are some of the very direct harms.
	\\&&
	(P3) ``now are there areas we are being wrong is more or less harmful, I think maybe based on the horizon on which we're forecasting, for example for macroeconomic, all these kind of things, 10 years, 20 years ahead, does that really change that much things. There will be so much things that's going to change anyway in between that the assumptions will be totally different, and it will appear to be wrong. And there are so many reasons why it's going to be wrong, but I think it's not going to be that much harmful because it's the … Maybe based on the horizon of the forecast it changed the – or maybe not the horizon, but the horizon of the decisions, there are some ways to – some alternatives or ways to change or re-impact the decisions if we were wrong. It's the ability to – you know, the agility, to say, `OK, we were wrong, but we could soften the impact.' We still have time to soften the impact."
	\\&&
	(P1) ``So how close is the wall? You know, you are running just – and you're going to hit the wall, how close is the wall, and do you have time to react and to turn left or right to avoid it or not? That could be a way to say OK, this domain is more likely to be harmful because if I'm wrong I have no Plan B. Whereas in other domains there might be some Plan B. For example, climate change, maybe 20 years or 30 years ago there were some – a lot of Plan B, and people tend to say, `OK, we're not going to make any decisions anyway,' so … And right now it seems like there's no – not that much Plan B, [laughs] and we are already close to the wall. And it's not a question of forecast anymore, by the way, it's a question of a decision, and it could harm. [Laughs].
	\\&&
	(P6) ``the harm was to the operations department who couldn’t cope. They were believing the forecasts, at least for a few weeks, and they were unable to cope with getting out the welcome packs to customers and connecting customers and so on. So the customers were suffering through delays because the forecasts were deliberately being kept too low."
	\\
	\bottomrule
\end{longtable}

\subsubsection*{Risk of Injury / Emotional or psychological injury /   Reduced self-esteem/reputation damage}
\vspace*{-0.4cm}
\begin{longtable}[l]{@{}ABC@{}}
	\toprule
	\textbf{Description in Miscrosoft}                          &
	\textbf{Description adapted to forecasting}                 &
	\textbf{Quotes from interviews}                         \\
	\midrule
	Some shared content can be harmful, false, misleading, or denigrating.   &
	Inaccurate forecasts could harm the reputation of individuals, businesses, or institutions that depend on them, potentially leading to diminished trust and credibility. &
	(P2) ``And if of course you are in the business of alarmism, and your livelihoods depends on getting donations from people who think like you are, then if these people donate less because they’re less alarmed nowadays now that they economists have published their forecasts, then this may also hurt you financially, and of course in terms of status and all kinds of other things that people rely on."
	\par
	(P5) ``There can be a reputational risk to being wrong if you put a forecast out there and say this is what we think is going to happen. I mean I still remember when in 2008 we had the global financial crisis. The former Fed Chairman, Alan Greenspan had a book that was still on bookstore shelves at the time that this happened, saying that he didn’t think there’d be another recession in 50 years. So there's a reputational cost of being wrong."
	\\&&
	(P10) ``Whenever your forecast is low and you're losing some demand, yes people tend to criticise that this is due to the forecast and ignoring the fact that this is supposed to be a stochastic type of signal and there's supposed to be error there."
	\\&&
	(P11) ``So I think the unspoken harm outside of the forecasting or demand planning departments is the fact that it's the most thankless job around if you're the one in charge of that forecast."
	\\&&
	(P13) ``I think another example is what I alluded to previously of forecasts that have come out under one set of conditions...Then they made decisions around setting up their production in line with those forecasts and then the circumstances changes and those volumes didn't materialise, and they then really felt somewhat betrayed by the donor community and by feeling like they'd done their part."
	\\&&
	(P18) ``But because of that 100\% accountability that can be created on forecasting, generally forecasting becomes like a Cinderella, right. So, we love Cinderella but at the end of the day, unfortunately, all the blame accumulates on forecasting. I think, mostly it leads to hurting trust in organisations within a big corporate world."
	\\&&
	(P17) ``It can make you look incompetent in the sense that they say to you, `We’re paying you all this money. Why can’t you give us a good forecast?' The answer, the scientific answer is, `Well, there’s this much level of uncertainty in what you’re asking me to forecast.' But if your boss is not happy with that you risk damage to your reputation, your career and getting fired for trying to give an honest answer to the organisation."
	\\&&
	(P4) ``also on the public health staff side, their perception, their reservations are predominantly based around, I guess, loss of trust – loss of public trust in that if they're making poor decisions or they have a lack of knowledge or so on. So for them it's more of a long term. They don’t have any immediate concern about a specific forecast, but yes, their concern I think is primarily a loss of trust with loss of public trust. If the forecasts are wrong or lead to poor decisions, for COVID-19 yes absolutely, we were involved in work and saying what do we think, when should we lift different restrictions during the extended lockdowns in Melbourne.
	\\&&
	(P16) ``But if you start to disseminate impact-based forecasts and they are not right, it can have a large impact on, yes, your trustworthiness. That’s a harm, not on the people that use it, but on yourself as an organisation, I guess.
	\\&&
	(P6) ``Damage to the forecasting industry as a result of these things? Well, there’s lots of disparaging jokes about forecasters. I’m sure you’ve heard a few of them. Taleb has said some forecasters cause more damage to society than criminals. He said that in the \emph{Black Swan}. And in 2014 when oil price forecasts were way out, Nigel Lawson in the Sunday times was totally disparaging forecasting as an industry and he said, `My yearend forecast, there’s no future for prediction.'”
	\\&&
	(P21) ``I definitely think that happened in COVID that collectively, as you accumulate more and more forecasts which are wrong because of course, they can't always be right because there's uncertainty. It's very easy to paint a picture of forecasts are useless, even though that's not a rational thing to conclude but it's a very easy story to tell."
	\\&&
	(P11) ``I’ve related to the fact that there’s no degree of uncertainty inherent in the forecast is that the consumers don’t realise or recognise that that number is uncertain and to what degree. And so if things go wrong, the forecaster or demand planner is always blamed"
	\\&&
	(P4) ``the public health staff, for better or worse, have a default mindset that making anything available publicly is opening themselves up to criticism and just negative impacts."
	\\&&
	(P6) ``The third thing is that I think the first two things can lead to damage to the forecasting profession. People say, `Oh, these forecasts are always wrong. These forecasts are inaccurate,' and this can lead people to dismiss forecasts when they’re reliable. All forecasts get tarred with the same brush. So there can be damage, I think, to forecasting in general when forecasting can be useful."
	\\&&
	(P5) ``There can be a reputational risk to being wrong if you put a forecast out there and say this is what we think is going to happen. I mean I still remember when in 2008 we had the global financial crisis. The former Fed Chairman, Alan Greenspan had a book that was still on bookstore shelves at the time that this happened, saying that he didn't think there'd be another recession in 50 years. So there’s a reputational cost of being wrong. And then I think on the being right side of the ledger, I think one potential risk that can come up if you're right, is that people might accuse you of having caused the thing to happen, or that there's this danger of making things a self-fulfilling prophecy."
	\\&&
	(P18) ``we basically plan for lots of inventory needs but then we don't necessarily see a demand and then that additional inventory, first it is a financial burden to the organisation. So much money is left. The trust forecasters have within the organisations get a hit as well. And this can happen in the other way as well."
	\\&&
	(P19) ``with the flu forecast in particular, especially early on in a season where there's not much of a signal, there's this quite a wide spread of uncertainty in the forecast predictions. And they – well, some people on the public health side would perceive making that available to the general public as an admission that they don't know what's going on. Even though the uncertainty is appropriate given the available data about how that might play out they see it as being something that a person could easily point at and say it's so uncertain to have – you know, our public health people don’t know what’s going on."
	\\&&
	(P20) ``yes, trust is possible. That's certainly one part of it. But, yes and losing trust is a long term negative consequence. And we've seen that and epidemiology in the US and they vaccinate very large under George W. Bush, vaccinate a large proportion of the population and the epidemic didn't develop at all."
	\\&&
	(P21) ``you can lose credibility but not through poor science but through poor inability, say to have effectively communicated it to the decision makers"
	\\
	\bottomrule
\end{longtable}

\subsubsection*{Risk of Injury / Emotional or psychological injury / Addiction/attention hijacking}
\vspace*{-0.4cm}
\begin{longtable}[l]{@{}ABC@{}}
	\toprule
	\textbf{Description in Miscrosoft}                                            &
	\textbf{Description adapted to forecasting}                                   &
	\textbf{Quotes from interviews}                                           \\
	\midrule
	Technology could be designed for prolonged interaction, without regard for well-being."   &
	Excessive dependence on short-term, frequent forecasts could lead to attention being drawn away from long-term strategic planning, causing reactive rather than proactive decision-making. &
	(P4) ``But yes, I guess any forecasts where your population’s perception of risk or harm to their self -- to themselves and, say, their friends and family, their immediate social networks, I guess I can just see it as a real potential for societal damage. [Laughs]." \par
	(P11) ``You ask any planner, they're always firefighting. They're going from crisis to crisis. That's not a happy place to be."
	\\
	\bottomrule
\end{longtable}

\subsubsection*{Risk of Injury / Emotional or psychological injury / Anxiety, stress, and emotional distress}
\vspace*{-0.4cm}
\begin{longtable}[l]{@{}ABC@{}}
	\toprule
	\textbf{Description in Miscrosoft}              &
	\textbf{Description adapted to forecasting}     &
	\textbf{Quotes from interviews}             \\
	\midrule
	&
	Psychological and emotional strain experienced by individuals or groups when facing uncertain or potentially adverse future outcomes predicted by forecasts. &
	(P4) ``And if you do it too early you've gone through all of the economic harm, the mental health and wellbeing harm, the impact on medical services beyond COVID-19 such as surgeries and so on, all of these wide-ranging societal impacts. And if you lift the lockdowns too early you then see the disease activity you were trying to prevent. So you've had all of the incidental harm of your intervention and you've lost any potential gain. So that was absolutely the most acute period where a real mistake in the forecast that said, yes, we should open, or yes we should relax some of the restrictions too early, but, yes, we would have gone through a very long, difficult time with very broad societal impact and then thrown away most if or if not all of what we would've hoped to achieved with that"
	\\&&
	(P8) ``for family planning I think the specific harm would be unplanned pregnancy because they don’t have their contraceptive methods in time"
	\\&&
	(P6) ``in terms of it’s best not to know, there’s a terrible disease you may know of called Huntington’s chorea. And it’s a terrible disease because if your parents had it, you’ve got a 50\% chance of getting it yourself and a 50\% chance of passing it onto your children. It’s a very life-limiting, terrible illness, but there is a test which will tell you for certain whether you’re going to have Huntington’s chorea, and it’s debatable whether people would want to know that. They might want to know it to decide whether to have children or not, but that aside they may not want to know it. And only 5-10\% in the USA decide – who may be in danger because their parents have the disease, only 5-10\% decide to take the test despite it being perfectly accurate giving them a perfectly accurate indication of whether they’ll develop it in the future. So those are some examples I think of potential harm. I’m sure we can go on with many more."
	\\&&
	(P6) ``Another damage, I think, is what I call it’s best not to know, and this can apply in healthcare situations where an inaccurate forecast, or even an accurate one, can cause a lot of anxiety and distress which may not be justified, or possibly could be avoided. Again, we can talk about that a bit later if you want. So for example an accurate forecast of an unavoidable illness that you suffer from can ruin your last few years of life if you like."
	\\&&
	(P2) ``When I go shopping in the evening and I can't find strawberries anymore, then yes, there is emotional distress here."
	\\
	\bottomrule
\end{longtable}

\newpage

\subsubsection*{Denial of consequential services / Opportunity loss / Insurance and benefit discrimination}
\vspace*{-0.4cm}
\begin{longtable}[l]{@{}ABC@{}}
	\toprule
	\textbf{Description in Miscrosoft}       &
	\textbf{Description adapted to forecasting}     &
	\textbf{Quotes from interviews}    \\
	\midrule
	This includes denying people insurance, social assistance, or access to a clinical trial due to biased standards.     &
	Forecasts used could unfairly deny coverage or benefits to certain groups based on biased or inaccurate future estimation. &
	(P3) ``say for example, mapping poverty or any type of aid allocation, you're trying to get some sort of aid to the most vulnerable people. Of course, your models are going to be imperfect, and just by being imperfect you could argue you're creating some indirect harm because you didn't perfectly identify the most vulnerable people, and so the aid was not perfectly used. You could argue that that’s harm''
	\\&&
	(P9) ``if a forecast is showing there’s not sufficient market or the alternative where a forecast might be overly optimistic in terms of what that market could be and then it doesn’t materialise. And then a manufacturer may lose interest in that market and no longer trust forecasts that come through or choose to no longer invest in something which could be – right, it could go either way in terms of that risk of over promising on a market opportunity and causing investment versus, or even – or not making that investment because it doesn’t seem like there’s the market potential"
	\\&&
	(P9) ``the harm is people not having access to life saving products. So that if you have an opportunity to bring a product to market, but forecasts are saying, OK. That’s actually not going to – instead of then identifying what are the challenges? Is it an affordability issue? Is it something we can work on to make this product which may have need or demand or promise, but we’re not seeing it? It’s then disallowing access to something which could be a life or health improving product."
	\\&&
	(P3) ``we try to identify regions that are in biggest need of help. If we get that wrong, so that through the inefficiencies can lead to harm that maybe if we miss the most vulnerable people and we send resources to people who are actually better off, then indirect through the failure to help the most vulnerable we are causing harm"
	\\&&
	(P3) ``we're trying to map population displacement. Again, we are doing this of course with the aim of facilitating aid and having a better humanitarian response. But you could imagine that other war parties including maybe even Russia wants to use this in a different way. If our model shows, or there's likely to be a large number of children in this particular part of the country, of course on one hand you could send aid there or you could send missiles there because you want to maximise the potential harm or outcome."
	\\
	\bottomrule
\end{longtable}

\subsubsection*{Denial of consequential services / Opportunity loss / Digital divide/technological discrimination}
\vspace*{-0.4cm}
\begin{longtable}[l]{@{}ABC@{}}
	\toprule
	\textbf{Description in Miscrosoft}   &
	\textbf{Description adapted to forecasting}  &
	\textbf{Quotes from interviews}  \\
	\midrule
	Disproportionate access to the benefits of technology may leave some people less informed or equipped to participate in society. & Access to high-quality  tools may be limited, leaving some stakeholders less informed or equipped to benefit from accurate forecasts, thus exacerbating inequalities. &
	(P21) ``if our data is on, say only a subset of the population about their infection experience, and then we forecast that and then we evaluate that and we say, `It’s doing well.' And then that drives a decision, that can definitely set up like socioeconomic or just marginalised communities might be excluded from the data because they’re harder to reach."
	\\
	\bottomrule
\end{longtable}

\newpage

\subsubsection*{Denial of consequential services / Opportunity loss / Loss of potential investment}
\vspace*{-0.4cm}
\begin{longtable}[l]{@{}ABC@{}}
	\toprule
	\textbf{Description in Miscrosoft}       &
	\textbf{Description adapted to forecasting}  &
	\textbf{Quotes from interviews}  \\
	\midrule
	 &
	refers to the missed opportunities for  investement when inaccurate forecasts lead businesses to overestimate or underestimate market demand. &
	(P9) ``if a manufacturer over produces a product and then the demand does not materialise based on what was seen in the forecast, they then lose – potentially. I mean, the risk goes wide. They could either lose confidence in the market and then not choose to invest in a product with public health promise or applicability in the future. Or, in some instances, what we’ve seen with smaller companies, it’s a risk to their viability in terms of even being in business and being able to provide products to especially a market that’s dependent on more affordable products that some of the less sophisticated, less well financed organisations produce.
	\\&&
	(P13) ``the harm is people not having access to life-saving products, right? So if you have an opportunity to bring a product to market, but forecasts are saying, `OK, that’s actually not going to …' instead of then identifying, you know, `What are the challenges? Is it an affordability issue? Is it something we can work on to make this product which may have need or demand or promise but we’re not seeing it?' it’s then disallowing access to something which could be a life or health-improving product."
	\\&&
	(P9) ``Because if a forecast is showing there’s not sufficient market or the alternative where a forecast might be overly optimistic in terms of what that market could be and then it doesn’t materialise. And then a manufacturer may lose interest in that market and no longer trust forecasts that come through or choose to no longer invest in something which could be – right, it could go either way in terms of that risk of over promising on a market opportunity and causing investment versus, or even – or not making that investment because it doesn’t seem like there’s the market potential"
	\\&&
	(P2) ``I'd like to go back one thing; here's one more thing where forecasting can cause harm. If the forecaster has just gone to the ISF and published his wonderful new forecasting method that improved the [mape? 00:21:30] by .2 per cent by building a huge giant boosted transformer neural recurrent network, and then somebody implements that, and it does indeed improve forecasts accuracy, but the business outcome is the same as before as with the exponential smoothing that we had before, then where's the harm.
	The harm is in all the opportunity costs of all the resources that we invested in building this wonderful shiny, giant complex forecasting method because that takes up processing power, storage space, data science expertise, \dots"
	\\&&
	(P18) ``under forecasting, it might be raining, we might need umbrellas at a retailer and then like there is none, there's a revenue loss over the lost sales."
	\\
	\bottomrule
\end{longtable}

\subsubsection*{Denial of consequential services / Economic loss / Economic exploitation}
\vspace*{-0.4cm}
\begin{longtable}[l]{@{}ABC@{}}
	\toprule
	\textbf{Description in Miscrosoft}     &
	\textbf{Description adapted to forecasting}    &
	\textbf{Quotes from interviews}  \\
	\midrule
	People might be compelled or misled to work on something that impacts their dignity or well-being.  &
	Inaccurate or biased forecasts might compel individuals or businesses to make detrimental economic decisions, such as overcommitting resources or accepting poor employment conditions based on flawed data. &
	(P14)  ``Let's construct the example of a criminal organisation who wants to forecast its things to do their criminal activities. The better the forecast, the worse the harm." \par
	(P18): ``But for the long-term investments, due to the nature of the business, think again about building data centres like we are buying [lens? 00:13:44], we are building buildings, right. So, if there is a real inaccuracy in our forecast, then we are talking about, again, big investments that could hurt the business."
	\\
	\bottomrule
\end{longtable}

\newpage

\subsubsection*{Denial of consequential services / Economic loss / Devaluation of individual expertise}
\vspace*{-0.4cm}
\begin{longtable}[l]{@{}ABC@{}}
	\toprule
	\textbf{Description in Miscrosoft}     &
	\textbf{Description adapted to forecasting}    &
	\textbf{Quotes from interviews}  \\
	\midrule
    Technology may supplant the use of paid human expertise or labor. &
	Automated forecasting systems may replace human expertise, leading to a devaluation of professional judgment and reducing opportunities for human oversight and intervention. &
	(P1)  ``Spending a lot of time forecasting things that don’t require anything more than a [unintelligible 00:33:22] forecast."
	\par
	(P2) ``I'm not saying that this is a consequence of bad forecasts but that can be a consequence of bad forecasts. And that is a type of harm. That was not the harm of throwing stuff away, that was the harm of locations closing, people not being able to shop there anymore, investors in the company losing their money, people working at the company losing their money. So this could all be a harm coming from bad forecasts."
	\\&&
	(P3) ``If we get that wrong, so that through the inefficiencies can lead to harm that maybe if we miss the most vulnerable people and we send resources to people who are actually better off, then indirect through the failure to help the most vulnerable we are causing harm."
	\\&&
	(P4) ``And if you lift the lockdowns too early you then see the disease activity you were trying to prevent. So you've had all of the incidental harm of your intervention and you've lost any potential gain. So that was absolutely the most acute period where a real mistake in the forecast that said, yes, we should open, or yes we should relax some of the restrictions too early, but, yes, we would have gone through a very long, difficult time with very broad societal impact and then thrown away most if or if not all of what we would’ve hoped to achieved with that."
	\\&&
	(P5) ``If you’re right, that’s great; if you’re wrong that means the emergency has happened somewhere else and all your stuff was in the wrong place and so there was that additional cost potentially in human suffering and loss of life in getting those resources to where they actually needed to be as opposed to where you thought they were going to be."
	\\&&
	(P6) ``Often the difference in accuracy of using a simple method and a highly sophisticated on is quite small and not of concern. I think forecasts need to be made independently of vested interest."
	\\&&
	(P7) ``As a consequence, there were huge issues with financial and environmental implications of all this excess apparel that had been built, never sold, and even though it had been cheap for us to make it compared to what we sold it for, it had to be burned or buried."
	\\&&
	(P7) ``A company like Zillow...got in the business of buying to flip properties a few years ago, and they didn't anticipate a downturn in the housing market and got stuck with a lot of properties that they had ended up overpaying for and had to sell at a loss."
	\\&&
	(P8) ``And the harm there could be we're ordering too much, so when we have too much inventory. Or we order too little and then we have too low inventory. And it is, it's similar to the manufacturer --"
	\\&&
	(P9) ``Forecasts were developed under one set of circumstances and one set of assumptions, and then they made decisions around setting up their production in line with those forecasts. And then the circumstances changes and those volumes didn't materialise."
	\\&&
	(10) ``What you end up to see is that you get more than what you sell. And then you need to go through markdown, and for some items some of them will be just waste because they are perishable."
	\\&&
	(11) ``If a company is not able to sell what they could have produced, then it harms their competitiveness and ultimately that could lead to job loss or stress or what have you."
	\\&&
	(12) ``They were, rightfully so, an unconstrained forecast. This is what we see in the market, elevated because of marketing activities and what we saw. All those things driving into a supply response was creating unnecessary inventory because they didn't have the manufacturing capabilities to respond to it."
	\\&&
	(13) ``Forecasts were developed under one set of circumstances and one set of assumptions. Then they made decisions around setting up their production in line with those forecasts and then the circumstances changes and those volumes didn't materialise, and they then really felt somewhat betrayed by the donor community and by feeling like they'd done their part."
	\\&&
	(15) ``Often, I think you try to use it, but if it's significant excess, then of course, to your point, expiry date comes into play. That's why we just often end up destroying it and that then leads to, as you mentioned also, environmental impact as well."
	\\&&
	(17) ``Some companies might actually have to then get to lay-offs and to really reconsider their business."
	\\&&
	(20) ``So if folks were to use our forecast potentially in an uninformed way, it could lead to real harm so we just overbuy on the wrong stuff. But the biggest concern definitely is economically."
\\
\bottomrule
\end{longtable}

\subsubsection*{Denial of consequential services / Economic loss / Misuse or misdirection of resources}
\vspace*{-0.4cm}
\begin{longtable}[l]{@{}ABC@{}}
	\toprule
	\textbf{Description in Miscrosoft}     &
	\textbf{Description adapted to forecasting}    &
	\textbf{Quotes from interviews}  \\
	\midrule
	&
	Forecasts can lead to unnecessary expenditures, wasted time, misallocation of resources, and misguided decision-making. &
	(P6) ``a major harm – the easiest one to identify is the misdirection of resources as a result of inaccuracy. And so if we can minimise the inaccuracy of forecasts we can minimis harm. So a lot of harm comes from inaccuracy in addition. So let’s take stage one which is identifying a decision that requires a forecast. Well, in many companies I’ve been in people confuse forecasts with decisions and they confuse them with many other things as well."
	\\&&
	(P6) ``Construction projects I’ve mentioned where there’s a great maldistribution of resources as a result of inaccurate forecasts."
	\\&&
	(P6) ``if a forecast is inaccurate, and I think much inaccuracy is avoidable, it can lead to a misdirection of resources. So for example if a forecast is based on inappropriate methods or if judgemental forecast is input into the process and they suffer from cognitive biases like optimism bias or indeed the planning fallacy. You can talk about that later if you like. Or particularly motivational biases like sandbagging where you deliberately keep your forecast low, so you’ll look good if that forecast is exceeded for example. If these forecasts are published there could be a misdirection of resources."
	\\&&
	(P6) ``Obviously weather forecasts being wrong, forecasting hurricanes when they don’t occur can create great inconvenience to people. Mass evacuations. Not forecasting when they do occur, I think New Orleans may have been an example of that during the Bush administration, can cause tremendous harm because people aren’t equipped or ready to evacuate when a hurricane appears."
	\\&&
	(P1) ``Spending a lot of time forecasting things that don't require anything more than a [unintelligible 00:33:22] forecast. I mean it's mostly being efficient in our activities and not waste our time, so basically chasing mice in my example. And it appears that in fact there's a lot of forecast where [unintelligible 00:33:47] forecast is, or a basic exponential smoothing over, whatever forecast is good enough to make the right decision – not to be accurate, but to make the right decision. Because when you make the decision there's a lot of things to account for, at least in supply chain again which is a very specific use. But if your forecast has two units errors, and when you buy your pack size is 50 units, I mean two units won't change your decision. And trying to reduce it from two to one [laughs] is just a waste of time"
	\\&&
	(P5) ``we're going to pre-position supplies there – if you're right, that's great; if you're wrong that means the emergency has happened somewhere else and all your stuff was in the wrong place and so there was that additional cost"
\\
\bottomrule
\end{longtable}

\subsubsection*{Denial of consequential services / Economic loss / Financial loss}
\vspace*{-0.4cm}
\begin{longtable}[l]{@{}ABC@{}}
	\toprule
	\textbf{Description in Miscrosoft}     &
	\textbf{Description adapted to forecasting}    &
	\textbf{Quotes from interviews}  \\
	\midrule
	&
	Highlights the significant financial loss businesses and users may face &
	(P2) ``I need to drive farther away to find a different store where I can buy my stuff. And that is simply an economical cost because you have to drive longer, I have to spend more on gas and stuff like that"
	\\&&
	(P9) ``I think another example is what I alluded to previously of forecasts that have come out under one set of conditions, and I’m thinking of a specific product and a small supplier or manufacturer in China who was developing a product. They worked really hard to get through their processes to get the right quality assurance for their product, to get it registered in markets.\par
	Forecasts were developed under one set of circumstances and one set of assumptions, and then they made decisions around setting up their production in line with those forecasts. And then the circumstances changes and those volumes didn’t materialise, and they then really felt somewhat betrayed by the donor community and by feeling like they’d done their part. So that’s more of the like trust, and then wanting to engage with this public health community moving forward and then feeling – I assume it had detrimental affects to their business as well."
	\\&&
	(P20) ``if folks were to use our forecast potentially in an uninformed way, it could lead to environmental harm so we just overbuy on the wrong stuff. But the biggest concern definitely is economically. Again, in the example that I have, in pricing, we forecast prices at all kinds of different discount levels. So if you want to use our forecasts, you have to know what the prices are going to be for it. You have to have a notion of the likelihood of a price and if you assumed a too opIntervieweeistic scenario where we discount too high or discount too low, then that could lead to a substantial financial harm for the company"
	\\&&
	(P4) ``if you do it too early you’ve gone through all of the economic harm, the mental health and wellbeing harm, the impact on medical services beyond COVID-19 such as surgeries and so on, all of these wide-ranging societal impacts. And if you lift the lockdowns too early you then see the disease activity you were trying to prevent. So you’ve had all of the incidental harm of your intervention and you’ve lost any potential gain. So that was absolutely the most acute period where a real mistake in the forecast that said, yes, we should open, or yes we should relax some of the restrictions too early, but, yes, we would have gone through a very long, difficult time with very broad societal impact and then thrown away most if or if not all of what we would’ve hoped to achieved with that"
	\\&&
	(P7) ``If you’re concerned about keeping your customers happy and fulfilling orders and you have a forecast that either underestimates what that demand is likely to be or you don’t make decisions to [unintelligible 00:14:03] a proper inventory, you risk that failing to fulfil orders. There’s potential of some loss of revenue there if there’s not a substitute product that they’re willing to buy. And potential long-term of losing customers who don’t find you as a trustworthy supplier for their demand."
	\\&&
	(P18) ``It is the economical loss again, yeah. Investments take place, we make purchases, we buy inventories and then because the forecasting is inaccurate, those investments are wasted and the company needs to take different actions to deplete that."
	\\&&
	(P9) ``Like thinking of manufacturers in China that have had issues with being told that they were – you know, they went through a process to get WHO prequalification, they’ve invested in this product, they’ve really worked to get to a price point that should be affordable. And then geared up their production and then invested in making those products available, and then if the demand doesn’t materialise that’s a risk to their business and a risk to the future of this product."
	\\&&
	(P1) ``Not making the right decisions which leads to operational issues which lead to economical impacts"
	\\&&
	(P18) ``some forecasting took place at Meta for infrastructure capacity planning and there was a lot of investment in a certain type of, say, computer, a machine, a server. And then the forecasts were not accurate and we did not necessarily realise those forecasts.
	So, the amount of investment that we had was exceeding the demand significantly. And then we realised that huge gap, what we had was a lot of machines lying around, basically cannot be leveraged in for some other purposes. We are talking about billions of investments. So those servers, even today, we couldn’t find a way to leverage them. Idle resources, the investment is done and still today we are looking for how to use it because there is no other way for Meta to use these type of machines, a very specific example.
	\\&&
	(P17) ``that’s the economic harm really, I think, of the forecast. If I summarise, although I went a bit into the extent of the whole process here, but then we would be missing sales or we would have huge inventory costs."
	\\&&
	(P20) ``the forecasts at very high discount levels are very inaccurate. It’s even hard to say if they’re inaccurate but the uncertainty is very large. So if you consume our forecasts, if you just take a point forecast from those very high discount regimes where we have a lot of uncertainty and you build, let’s say, an inventory control mechanism for it, and this actually happened, then that can lead to substantial harm, financial loss."
	\\&&
	(P17) ``They kept on forecasting high, high, high and then the demand wasn’t there anymore. There are some companies that have massive write-offs because their inventory is huge, and after expiry date, well what can they do? They’re not going to sell it anymore. Those write-offs, if they’re big enough, can really, really cause big harm economically for companies, and some companies might actually have to then get to lay-offs and to really reconsider their business."
	\\&&
	(P2) ``they went out of business because – or partly because – they had issues with their on shelf availability, and people were not able to buy their stuff. I’m not saying that this is a consequence of bad forecasts but that can be a consequence of bad forecasts. And that is a type of harm. That was not the harm of throwing stuff away, that was the harm of locations closing, people not being able to shop there anymore, investors in the company losing their money, people working at the company losing their money. So this could all be a harm coming from bad forecasts."
	\\&&
	(P11) ``think of Shapley or what have you for machine learning, right, so that you can get an idea of what happened. What kind of input would have been driving certain outputs? If you don’t have that, it’s going to be very risky to implement it, because at some point in time you’re going to get that one blow up and not know why it happened and if it’s going to happen again."
	\\&&
	(P18) `` we basically plan for lots of inventory needs but then we don't necessarily see a demand and then that additional inventory, first it is a financial burden to the organisation. So much money is left."
	\\&&
	(P12) `` without the proper context, people can overreact or underreact to that signal, is what’s going to happen. With any overreaction or underreaction, you have the capability of not servicing the customer, which could have devastating consequences to it, or you could overreact and end up with way too much inventory. You’re going to have cost and cash issues, potential issues with those that could impact your margin, it can impact cashflow"
	\\
\bottomrule
\end{longtable}

\subsubsection*{Infringement on human rights / Privacy loss / Interference with private life}
\vspace*{-0.4cm}
\begin{longtable}[l]{@{}ABC@{}}
	\toprule
	\textbf{Description in Miscrosoft}     &
	\textbf{Description adapted to forecasting}    &
	\textbf{Quotes from interviews}  \\
	\midrule
	Revealing information a person has not chosen to share &
	Forecasting models that misuse personal data could reveal sensitive information about individuals, leading to breaches of privacy or unwanted exposure. &
	(P3) ``Again, it was collected with the best of intentions, but now the same data, especially because it's at the individual data level, is arguably being used for other purposes. Maybe there are even documented cases of people being persecuted or killed or abducted because of this data."
	\\
\bottomrule
\end{longtable}

\newpage

\subsubsection*{Infringement on human rights / Privacy loss / Loss of freedom of movement or assembly}
\vspace*{-0.4cm}
\begin{longtable}[l]{@{}ABC@{}}
	\toprule
	\textbf{Description in Miscrosoft}     &
	\textbf{Description adapted to forecasting}    &
	\textbf{Quotes from interviews}  \\
	\midrule
	This means an inability to navigate the physical or virtual world with desired anonymity. &
	Forecasts predicting certain trends may limit mobility or opportunities for specific groups, either through actual policy changes or self-imposed limitations based on expected outcomes. &
	(P3) ``European Asylum Search, no, asylum support organisation – EASO – that has also published this work on forecasting migration. And even their name suggests asylum support, they are a bit closer to Frontex like organisations than, let's say to humanitarian support organisations. And so of course we don't know exactly what they're doing behind the scenes, but I mean depending on what constituents you're representing, you could potentially argue that maybe there is some harm done there because maybe their forecasts are meant more in the sense of can we avoid those flows, can we do something to keep people in their country of origin, or somewhere in Africa where maybe you could argue that causes harm."
	\\&&
	(P5) ``you can't just publish that because if they see that they're going to think oh no, this means there's going to be protests coming up, there's going to be instability. We need to make sure we crack down really fast to keep that from happening. And so by publishing a prediction you are causing the thing that you are predicting to happen, or at least contributing to it in some sense. So I think that's a potential risk of being right.
	\\
	\bottomrule
\end{longtable}

\subsubsection*{Infringement on human rights / Environmental impact / Exploitation or depletion of resources}
\vspace*{-0.4cm}
\begin{longtable}[l]{@{}ABC@{}}
	\toprule
	\textbf{Description in Miscrosoft}     &
	\textbf{Description adapted to forecasting}    &
	\textbf{Quotes from interviews}  \\
	\midrule
	Obtaining the raw materials for technology, including how it's powered, leads to negative consequences to the environment and its inhabitants. &
	Forecasts predicting future resource needs might encourage unsustainable practices, such as over-extraction or depletion of environmental resources, based on inaccurate demand models. &
	(P2) ``And of course, conversely, perhaps the reaction to not having our forecast would be to have much higher safety stock. And actually I’m getting confused, but you know what I mean. So actually this person, this five per cent guy that’s somebody who was harmed by the correct forecast, and if you don’t have a correct forecast, what’s the answer then."
	\par
	(P6) ``I think the environmental impact is quite a powerful one, and I've argued -- my nephew used to say, `What's the point of all this research you do in forecasting?' and I said, `Well, if you can get more accurate forecasts there will be less waste, less things going to landfill and so on.'
	\\&&
	(P7) ``There were huge issues with financial and environmental implications of all this excess apparel that had been built, never sold, and even though it had been cheap for us to make it compared to what we sold it for, it had to be burned or buried."
	\\&&
	(P9) ``When products are over forecast and over procured, there is then the whole issue of waste disposal. How do you destroy products? What are the -- not just the products themselves and the active pharmaceutical ingredients, but also the packaging and everything that goes along with that."
	\\&&
	(P10) ``For the perishable items will be kind of some environmental impact because you are throwing something out."
	\\&&
	(P20) ``So if folks were to use our forecast potentially in an uninformed way, it could lead to real harm so we just overbuy on the wrong stuff."
	\\
\bottomrule
\end{longtable}

\newpage

\subsubsection*{Infringement on human rights / Environmental impact / (Electronic) waste}
\vspace*{-0.4cm}
\begin{longtable}[l]{@{}ABC@{}}
	\toprule
	\textbf{Description in Miscrosoft}     &
	\textbf{Description adapted to forecasting}    &
	\textbf{Quotes from interviews}  \\
	\midrule
	Reduced quality of collective well-being because of the inability to repair, recycle, or otherwise responsibly dispose of electronics. &
	Overreliance on forecasts encouraging constant technological upgrades may contribute to electronic waste if old systems are frequently discarded in favor of newer models. &
	(P13) ``There is then the whole issue of waste disposal, right? How do you destroy products? You know, not just the product themselves and the active pharmaceutical ingredients, but also the packaging and everything that goes along with that. That's certainly harm that we've seen done in terms of needing to destroy large quantities of products where they have expired or not been used."
	\par
	(P15) ``Often, I think you try to use it, but if it's significant excess, then of course, to your point, expiry date comes into play. That's why we just often end up destroying it and that then leads to, as you mentioned also, environmental impact as well."
	\\&&
	(P9) `` When products are over forecast and over procured, there is then the whole issue of waste disposal. How do you destroy products? What are the – not just the products themselves and the active pharmaceutical ingredients, but also the packaging and everything that goes along with that. And that’s certainly harm that we’ve seen done in terms of needing to destroy large quantities of products where they have expired or not been used."
	\\&&
	(P10) `` when people they are not satisfied with the result of forecast and they are trying to change it, typically it’s in the direction of increasing the forecast, magnitude of the forecast, because they are more worried about going out of stock. So as a result, when you are increasing your purchases, you’re kind of just dialling up your demand, overwriting it, what you end up to see is that you get more than what you sell. And then you need to go through markdown, and for some items some of them will be just waste because they are perishable."
	\\&&
	(P13) `` we’ve seen this too in working with national governments when products are over-forecast and over-procured. There is then the whole issue of waste disposal, right? How do you destroy products? You know, not just the product themselves and the active pharmaceutical ingredients, but also the packaging and everything that goes along with that. That’s certainly harm that we’ve seen done in terms of needing to destroy large quantities of products where they have expired or not been used."
	\\&&
	(P7) `` The other side, you know, overstocking your inventory, of course the cost, the potential waste of that. Disposal cost, overage."
	\\&&
	(P6) `` So inaccuracy or indeed adjustments caused by being over-optimistic, or political adjustments, inaccuracy caused by the wrong type of methods being used can I think do tremendous harm through all the waste"
	\\&&
	(P1) `` It could be harm because you’re actually somehow wasting the resource of the organisation, right."
	\\&&
	(P2) `` If people had put in too much safety stock then they might be throwing away, at the end of day, lots of food, fruit, vegetables."
	\\&&
	(P9) `` I mean, it’s harm to the environment, misuse of resources, waste of resources, waste of time, waste of distributing those products to places where they sit on shelves and don’t get used."
	\\
\bottomrule
\end{longtable}

\subsubsection*{Erosion of social \& democratic structures / Manipulation / Misinformation}
\vspace*{-0.4cm}
\begin{longtable}[l]{@{}ABC@{}}
	\toprule
	\textbf{Description in Miscrosoft}     &
	\textbf{Description adapted to forecasting}    &
	\textbf{Quotes from interviews}  \\
	\midrule
	Disguising fake information as legitimate or credible information &
	Forecasts based on incomplete or biased data could spread misinformation, influencing decisions in harmful ways by providing a false sense of accuracy or legitimacy. &
	(P5) ``I think one potential risk that can come up if you’re right, is that people might accuse you of having caused the thing to happen, or that there’s this danger of making things a self-fulfilling prophecy."
	\par
	(P12) ``There's a lot of forecast that happen that are used for political reasons to sway decisions, to sway policy, to sway people's opinion. That, I believe, is a harm and that's done intentionally by the people using the stats in some type of forecasting process, you know, a method that they -- someone who's just consuming that without any thought at the end, the general public can do damage, by all means, yes."
	\\&&
	(P13) ``Especially people who don't do forecasting work but really want the output. They just want the number without the context of what that might mean or the range of uncertainty or where there might be need to update and improve and not take that as gospel."
	\\&&
	(P16) ``Well, people could have been evacuated from the coastal areas and be brought to safe areas. I think that the seriousness of this storm surge was not well understood. So the actions that were taken were not strong enough, were not adequate."
	\\&&
	(P17) ``But if you're using basically anything that relates to PII in forecasting, there's definitely a risk of harm by carrying over biases which can be based on gender, which can be based on ethnical belonging, which can be on a lot of different things, right?"
	\\&&
	(P19) ``It's the interpretation of the forecast as a fact. That’s a future fact, if there is such a thing as a future fact."
	\\&&
	(P21) ``And they spoke publicly about that. They didn’t show the forecast because they were all kept in secret boxes. But they said essentially, `The scientists have told us it’s going to be terrible' with no nuance or layering of that. And it had the risk, I think, in the media of causing some panic."
	\\&&
	(P12) ``there’s a lot of forecast that happen that are used for political reasons to sway decisions, to sway policy, to sway people’s opinion. That, I believe, is a harm and that’s done intentionally by the people using the stats in some type of forecasting process, you know, a method that they – someone who’s just consuming that without any thought at the end, the general public can do damage,"
	\\
	\bottomrule
\end{longtable}

\subsubsection*{Erosion of social \& democratic structures / Manipulation / Behavioral exploitation}
\vspace*{-0.4cm}
\begin{longtable}[l]{@{}ABC@{}}
	\toprule
	\textbf{Description in Miscrosoft}     &
	\textbf{Description adapted to forecasting}    &
	\textbf{Quotes from interviews}  \\
	\midrule
	This means exploiting personal preferences or patterns of behavior to induce a desired reaction. &
	Forecasts based on behavioral patterns might exploit personal habits, nudging stakeholders into decisions that align with the model's predictions rather than their actual best interests. &
	(P4) ``So I am aware of people who use any messaging, be it forecasts or just random voices in the media or, yes, media coverage of a single high profile thing such as, say, a child being hospitalised or put into an intensive care unit. They will use that to drum up or to help sell the message of you should get vaccinated. And I am aware of a few people in this space that push vaccination and see anything that supports the idea that there’s going to be a lot of disease is an acceptable means of trying to convince more of your population to get vaccinated regardless of how ill-founded or incorrect, say, the forecasts or the data or your messages are."
	\\&&
	(P6) ``The fourth thing, I think, is what I call the self-destructive forecast. In many domains, when you make a forecast it actually changes the thing you're trying to forecast."
	\\&&
	(P6) ``The fourth thing, I think, is what I call the self-destructive forecast. In many domains, when you make a forecast it actually changes the thing you’re trying to forecast. I mean, a classic thing is opinion polls where you forecast that one party is going to win an election and voters decide not to bother voting, `Oh, well they’re going to win by a landslide. There’s no point in turning out.' They don’t vote. The forecast turns out to be wrong, and this happened in the 1970 election, well before you were born I’m sure, in Britian where Harold Wilson’s Labour government were looking for a landslide. Lovely summer’s day, Labour voters didn’t bother to vote, and Edward Heath won the election. I think because of this, opinion poll forecasts have been banned apparently in lots of different countries because of their danger of influencing the electorate, and possibly damaging democracy. So the damage can be to democracy."
	\\&&
	(P6) ``Self-destructive forecasts we’ve mentioned. Opinion polls which arguably damage democracy if people’s voting intentions are changed as a result of those polls."
	\\&&
	(P6) ``Opinion polls which arguably damage democracy if people’s voting intentions are changed as a result of those polls."
	\\&&
	(P19) ``I haven't mentioned election forecasting. It's a timely reminder that people can then choose not to vote because they think they can. And Brexit is sometimes used as an example. I don't know whether it is an example or not, but sometimes used, that people believe that the vote for Brexit, against Brexit was straightforwardly won and didn't vote and therefore, I don't know. There may be some evidence about that.
	So, yes, there are clearly potential negative consequences from trusting, putting 100
	\\&&
	(P1) ``in France, for example, people tend – there's very few people going to vote. Most people are not doing that, or they don't care, and/or they vote blank, not to … And so in fact there's very few people that make the decisions. So a little change; it was very few people could change the final outcome. And when the vote forecast are published, sometimes they say, `Oh it's very, very close and we don't know who's going to win,' so basically we – that's – it doesn't change the way I think people act, or maybe they could change the way they act. But sometimes they say, `OK, this candidate, this guy, is going to win with a very large advance.'
	And then people in his electoral side say, `OK, why should I go voting? I'd rather go to the restaurant, or go on vacation, or whatsoever, but I'm not going to vote because it won't change anything. It’s already a win.' So people tend to think that OK, everything is already OK, clear. I could basically discard this, making any decision, it's already done. And because of that, the candidate is losing because a lot of people made this decision, this assumption, and it changed the way people acted. Or maybe the candidate himself changed his behaviour and said, `OK, it's going to be … ' And because of that, he feels more relaxed [laughs] and is not fighting up to the end."
	\\
	\bottomrule
\end{longtable}

\subsubsection*{Erosion of social \& democratic structures / Social detriment / Stereotype reinforcement}
\vspace*{-0.4cm}
\begin{longtable}[l]{@{}ABC@{}}
	\toprule
	\textbf{Description in Miscrosoft}     &
	\textbf{Description adapted to forecasting}    &
	\textbf{Quotes from interviews}  \\
	\midrule
	This may perpetuate uninformed `conventional wisdom' about historically or statistically underrepresented people. &
	Forecasts that rely on historical data could perpetuate harmful stereotypes by reinforcing existing biases, particularly in areas like employment, education, or criminal justice." &
	(P21) ``if our data is on, say only a subset of the population about their infection experience, and then we forecast that and then we evaluate that and we say, `It’s doing well.' And then that drives a decision, that can definitely set up like socioeconomic or just marginalised communities might be excluded from the data because they’re harder to reach."
	\\
	\bottomrule
\end{longtable}

\subsubsection*{Erosion of social \& democratic structures / Social detriment / Loss of representation}
\vspace*{-0.4cm}
\begin{longtable}[l]{@{}ABC@{}}
	\toprule
	\textbf{Description in Miscrosoft}     &
	\textbf{Description adapted to forecasting}    &
	\textbf{Quotes from interviews}  \\
	\midrule
	Broad categories of generalization obscure, diminish, or erase real identities. &
	Broad generalizations in forecasting models may obscure the needs of minority or underrepresented groups, leading to decisions that fail to account for their specific needs or circumstances. &
	(P5) ``I think one example where I was looking at World Bank data and showing that inequality was trending downward, and everybody in the audience was up in arms saying, `That can't possibly be true. Inequality has to be going up; it's the worst that it's ever been.' And they live there, and I don't live there; I couldn't really argue with them too much on that other than to say the official statistics are what they are, but if no one believes those statistics in the first place it's a really hard conversation to have."
	(P21) ``if our data is on, say only a subset of the population about their infection experience, and then we forecast that and then we evaluate that and we say, `It’s doing well.' And then that drives a decision, that can definitely set up like socioeconomic or just marginalised communities might be excluded from the data because they’re harder to reach."
	\\
	\bottomrule
\end{longtable}

\end{document}

%% file: harm_matrix.tex
\begin{tikzpicture}[x=1pt,y=1pt]
\definecolor{fillColor}{RGB}{255,255,255}
\path[use as bounding box,fill=fillColor,fill opacity=0.00] (0,0) rectangle (289.08,289.08);
\begin{scope}
\path[clip] (  0.00,  0.00) rectangle (289.08,289.08);
\definecolor{drawColor}{RGB}{0,0,0}

\path[draw=drawColor,line width= 0.4pt,line join=round,line cap=round] ( 49.20, 61.20) --
	(263.88, 61.20) --
	(263.88,239.88) --
	( 49.20,239.88) --
	cycle;
\end{scope}
\begin{scope}
\path[clip] (  0.00,  0.00) rectangle (289.08,289.08);
\definecolor{drawColor}{RGB}{0,0,0}

\node[text=drawColor,anchor=base west,inner sep=0pt, outer sep=0pt, scale=  1.00] at (110.52, 17.54) {\bfseries Forecast Accuracy};

\node[text=drawColor,rotate= 90.00,anchor=base west,inner sep=0pt, outer sep=0pt, scale=  1.00] at (  9.83,118.82) {\bfseries Harm Intent};
\end{scope}
\begin{scope}
\path[clip] ( 49.20, 61.20) rectangle (263.88,239.88);
\definecolor{drawColor}{RGB}{0,0,0}

\path[draw=drawColor,line width= 0.4pt,dash pattern=on 4pt off 4pt ,line join=round,line cap=round] ( 49.20,150.54) -- (263.88,150.54);

\path[draw=drawColor,line width= 0.4pt,dash pattern=on 4pt off 4pt ,line join=round,line cap=round] (156.54, 61.20) -- (156.54,239.88);
\end{scope}
\begin{scope}
\path[clip] (  0.00,  0.00) rectangle (289.08,289.08);
\definecolor{drawColor}{RGB}{0,0,0}

\node[text=drawColor,anchor=base west,inner sep=0pt, outer sep=0pt, scale=  1.00] at ( 67.20, 51.60) {Accurate};

\node[text=drawColor,anchor=base east,inner sep=0pt, outer sep=0pt, scale=  1.00] at (245.88, 51.60) {Inaccurate};

\node[text=drawColor,rotate= 90.00,anchor=base west,inner sep=0pt, outer sep=0pt, scale=  1.00] at ( 46.80, 67.82) {Intentional};

\node[text=drawColor,rotate= 90.00,anchor=base east,inner sep=0pt, outer sep=0pt, scale=  1.00] at ( 46.80,233.26) {Unintentional};
\end{scope}
\begin{scope}
\path[clip] ( 49.20, 61.20) rectangle (263.88,239.88);
\definecolor{drawColor}{RGB}{0,0,0}

\node[text=drawColor,anchor=base,inner sep=0pt, outer sep=0pt, scale=  1.00] at (102.94,197.77) {Unintentional harm};

\node[text=drawColor,anchor=base,inner sep=0pt, outer sep=0pt, scale=  1.00] at (102.94,185.77) { accurate forecasts};

\node[text=drawColor,anchor=base,inner sep=0pt, outer sep=0pt, scale=  1.00] at (210.14,197.77) {Unintentional harm};

\node[text=drawColor,anchor=base,inner sep=0pt, outer sep=0pt, scale=  1.00] at (210.14,185.77) { inaccurate forecasts};

\node[text=drawColor,anchor=base,inner sep=0pt, outer sep=0pt, scale=  1.00] at (102.94,115.04) {Intentional harm};

\node[text=drawColor,anchor=base,inner sep=0pt, outer sep=0pt, scale=  1.00] at (102.94,103.04) { accurate forecasts};

\node[text=drawColor,anchor=base,inner sep=0pt, outer sep=0pt, scale=  1.00] at (210.14,115.04) {Intentional harm};

\node[text=drawColor,anchor=base,inner sep=0pt, outer sep=0pt, scale=  1.00] at (210.14,103.04) { inaccurate forecasts};
\end{scope}
\end{tikzpicture}

%% file: literature.bib
@book{montgomeryTimeSeriesBook2015,
    author = {Montgomery, DC and Jennings, CL and Kulahci, M},
    title = {Introduction to time series analysis and forecasting},
    publisher ={John Wiley and Sons},
    year = {2015}
}

@book{thompson2022escape,
  title={Escape from model land: How mathematical models can lead us astray and what we can do about it},
  author={Thompson, Erica},
  year={2022},
  publisher={Hachette UK}
}

@article{aizenberg2020designing,
  title     = {Designing for human rights in {AI}},
  author    = {Aizenberg, Evgeni and Van Den Hoven, Jeroen},
  journal   = {Big Data \& Society},
  volume    = {7},
  number    = {2},
  pages     = {2053951720949566},
  year      = {2020},
  publisher = {SAGE Publications},
}

@article{arrieta2020explainable,
  title     = {Explainable Artificial Intelligence ({XAI}): Concepts, taxonomies, opportunities and challenges toward responsible {AI}},
  author    = {Arrieta, Alejandro Barredo and D{\'\i}az-Rodr{\'\i}guez, Natalia and Del Ser, Javier and Bennetot, Adrien and Tabik, Siham and Barbado, Alberto and Garc{\'\i}a, Salvador and Gil-L{\'o}pez, Sergio and Molina, Daniel and Benjamins, Richard and others},
  journal   = {Information Fusion},
  volume    = {58},
  pages     = {82--115},
  year      = {2020},
  publisher = {Elsevier}
}

@book{asimov1978bicentennial,
  title     = {The bicentennial man},
  author    = {Asimov, Isaac},
  publisher = {Panther},
  year      = {1978}
}

@article{bandura2002selective,
  title     = {Selective moral disengagement in the exercise of moral agency},
  author    = {Bandura, Albert},
  journal   = {Journal of Moral Education},
  volume    = {31},
  number    = {2},
  pages     = {101--119},
  year      = {2002},
  publisher = {Taylor \& Francis}
}

@book{barocas-hardt-narayanan,
  title     = {Fairness and machine learning: Limitations and opportunities},
  author    = {Solon Barocas and Moritz Hardt and Arvind Narayanan},
  url       = {http://www.fairmlbook.org},
  publisher = {MIT Press},
  year      = {2023}
}

@inproceedings{barocas2017problem,
  title     = {The problem with bias: from allocative to representational harms in machine learning},
  booktitle = {9th annual {SIGCIS} conference},
  author    = {Barocas, Solon and Crawford, Kate and Shapiro, Aaron and Wallach, Hanna},
  year      = {2017},
  address   = {Philadelphia, PA, USA}
}

@book{beauchamp2001principles,
  title={Principles of biomedical ethics},
  author={Beauchamp, Tom L and Childress, James F},
  year={2001},
  publisher={Oxford University Press},
  address = {USA}
}

@article{binns2022human,
  title     = {Human Judgment in algorithmic loops: Individual justice and automated decision-making},
  author    = {Binns, Reuben},
  journal   = {Regulation \& Governance},
  volume    = {16},
  number    = {1},
  pages     = {197--211},
  year      = {2022},
  publisher = {Wiley Online Library}
}

@article{bradley2012doing,
  title     = {Doing away with harm},
  author    = {Bradley, Ben},
  journal   = {Philosophy and Phenomenological Research},
  volume    = {85},
  number    = {2},
  pages     = {390--412},
  year      = {2012},
  publisher = {Wiley Online Library}
}

@book{cane2002responsibility,
  title     = {Responsibility in law and morality},
  author    = {Cane, Peter},
  year      = {2002},
  publisher = {Bloomsbury Publishing}
}

@inproceedings{chan2023harms,
  title     = {Harms from Increasingly Agentic Algorithmic Systems},
  author    = {Chan, Alan and Salganik, Rebecca and Markelius, Alva and Pang, Chris and Rajkumar, Nitarshan and Krasheninnikov, Dmitrii and Langosco, Lauro and He, Zhonghao and Duan, Yawen and Carroll, Micah and others},
  booktitle = {Proceedings of the 2023 Conference on Fairness, Accountability, and Transparency},
  pages     = {651--666},
  year      = {2023},
  publisher = {Association for Computing Machinery}
}

@book{crawford2021atlas,
  title     = {The atlas of {AI}: Power, politics, and the planetary costs of artificial intelligence},
  author    = {Crawford, Kate},
  year      = {2021},
  publisher = {Yale University Press}
}

@incollection{davydenko2016forecast,
  title     = {Forecast error measures: critical review and practical recommendations},
  author    = {Davydenko, Andrey and Fildes, Robert},
  booktitle   = {Business forecasting: Practical problems and solutions},
  year      = {2016},
  publisher = {John Wiley \& Sons}
}

@book{dignum2019responsible,
  title     = {Responsible artificial intelligence: How to develop and use {AI} in a responsible way},
  author    = {Dignum, Virginia},
  year      = {2019},
  publisher = {Springer}
}

@book{feinberg1984harm,
  title     = {Harm to others},
  author    = {Feinberg, Joel},
  year      = {1984},
  publisher = {Oxford University Press}
}

@article{finkelstein2002risk,
  title     = {Is risk a harm},
  author    = {Finkelstein, Claire},
  journal   = {University of Pennsylvania Law Review},
  volume    = {151},
  pages     = {963},
  year      = {2003},
  publisher = {HeinOnline}
}

@article{fischhoff2014communicating,
  title     = {Communicating scientific uncertainty},
  author    = {Fischhoff, Baruch and Davis, Alex L},
  journal   = {Proceedings of the National Academy of Sciences},
  volume    = {111},
  number    = {supplement\_4},
  pages     = {13664--13671},
  year      = {2014},
  publisher = {National Acad Sciences}
}

@article{foot1967problem,
  title   = {The problem of abortion and the doctrine of the double effect},
  author  = {Foot, Philippa},
  year    = {1967},
  journal = {Oxford Review},
  number  = 5
}

@article{goodsocietyfloridi2021ethical,
  title     = {An ethical framework for a good {AI} society: Opportunities, risks, principles, and recommendations},
  author    = {Floridi, Luciano and Cowls, Josh and Beltrametti, Monica and Chatila, Raja and Chazerand, Patrice and Dignum, Virginia and Luetge, Christoph and Madelin, Robert and Pagallo, Ugo and Rossi, Francesca and others},
  journal   = {Ethics, governance, and policies in artificial intelligence},
  pages     = {19--39},
  year      = {2021},
  publisher = {Springer}
}

@article{greene2019adjusting,
  title     = {Adjusting to the {GDPR}: The impact on data scientists and behavioral researchers},
  author    = {Greene, Travis and Shmueli, Galit and Ray, Soumya and Fell, Jan},
  journal   = {Big data},
  volume    = {7},
  number    = {3},
  pages     = {140--162},
  year      = {2019},
  publisher = {Mary Ann Liebert, Inc., publishers 140 Huguenot Street, 3rd Floor New~…}
}

@article{hardt2016equality,
  title   = {Equality of opportunity in supervised learning},
  author  = {Hardt, Moritz and Price, Eric and Srebro, Nati},
  journal = {Advances in neural information processing systems},
  volume  = {29},
  year    = {2016}
}

@incollection{harman2009harming,
  title     = {Harming as causing harm},
  author    = {Harman, Elizabeth},
  booktitle = {Harming future persons: Ethics, genetics and the nonidentity problem},
  editor    = {Roberts, M and Wasserman, D},
  pages     = {137--154},
  year      = {2009},
  publisher = {Springer}
}

@article{hayenhjelm2012moral,
  title     = {The moral problem of risk impositions: A survey of the literature},
  author    = {Hayenhjelm, Madeleine and Wolff, Jonathan},
  journal   = {European Journal of Philosophy},
  volume    = {20},
  pages     = {E26--E51},
  year      = {2012},
  publisher = {Wiley Online Library}
}

@article{hewamalage2021recurrent,
  title     = {Recurrent neural networks for time series forecasting: Current status and future directions},
  author    = {Hewamalage, Hansika and Bergmeir, Christoph and Bandara, Kasun},
  journal   = {International Journal of Forecasting},
  volume    = {37},
  number    = {1},
  pages     = {388--427},
  year      = {2021},
  publisher = {Elsevier}
}

@article{hyams2013ethics,
  title     = {The ethics of carbon offsetting},
  author    = {Hyams, Keith and Fawcett, Tina},
  journal   = {Wiley Interdisciplinary Reviews: Climate Change},
  volume    = {4},
  number    = {2},
  pages     = {91--98},
  year      = {2013},
  publisher = {Wiley Online Library}
}

@article{hyndman2006another,
  title     = {Another look at measures of forecast accuracy},
  author    = {Hyndman, Rob J and Koehler, Anne B},
  journal   = {International Journal of Forecasting},
  volume    = {22},
  number    = {4},
  pages     = {679--688},
  year      = {2006},
  publisher = {Elsevier}
}

@book{hyndman2018forecasting,
  title     = {Forecasting: Principles and practice},
  author    = {Hyndman, Rob J and Athanasopoulos, George},
  year      = {2021},
  publisher = {OTexts},
  edition = {3rd}
}

@article{karliuk2023proportionality,
  title     = {Proportionality principle for the ethics of artificial intelligence},
  author    = {Karliuk, Maksim},
  journal   = {AI and Ethics},
  volume    = {3},
  number    = {3},
  pages     = {985--990},
  year      = {2023},
  publisher = {Springer}
}

@article{king2023self,
  title     = {Self-fulfilling Prophecy in Practical and Automated Prediction},
  author    = {King, Owen C and Mertens, Mayli},
  journal   = {Ethical Theory and Moral Practice},
  pages     = {1--26},
  year      = {2023},
  publisher = {Springer}
}

@article{klocksiem2012defense,
  title     = {A defense of the counterfactual comparative account of harm},
  author    = {Klocksiem, Justin},
  journal   = {American Philosophical Quarterly},
  volume    = {49},
  number    = {4},
  pages     = {285--300},
  year      = {2012},
  publisher = {JSTOR}
}

@article{koepke2018danger,
  title     = {Danger ahead: Risk assessment and the future of bail reform},
  author    = {Koepke, John Logan and Robinson, David G},
  journal   = {Washington Law Review},
  volume    = {93},
  pages     = {1725},
  year      = {2018},
  publisher = {HeinOnline}
}

@article{layton2021fighting,
  title   = {Fighting Artificial Intelligence Battles: Operational Concepts for Future {AI}-Enabled Wars},
  author  = {Layton, Peter},
  journal = {Network},
  volume  = {4},
  number  = {20},
  pages   = {1--100},
  year    = {2021}
}

@article{makridakis2020social,
  title     = {Forecasting in social settings: The state of the art},
  author    = {Makridakis, Spyros and Hyndman, Rob J and Petropoulos, Fotios},
  journal   = {International Journal of Forecasting},
  volume    = {36},
  number    = {1},
  pages     = {15--28},
  year      = {2020},
  publisher = {Elsevier}
}

@inproceedings{mitchell2019model,
  title     = {Model cards for model reporting},
  author    = {Mitchell, Margaret and Wu, Simone and Zaldivar, Andrew and Barnes, Parker and Vasserman, Lucy and Hutchinson, Ben and Spitzer, Elena and Raji, Inioluwa Deborah and Gebru, Timnit},
  booktitle = {Proceedings of the 2019 Conference on Fairness, Accountability, and Transparency},
  pages     = {220--229},
  year      = {2019},
  publisher = {Association for Computing Machinery}
}

@misc{msharms,
  title  = {Types of harm},
  author = {Microsoft},
  year   = {2023},
  url    = {https://learn.microsoft.com/en-us/azure/architecture/guide/responsible-innovation/harms-modeling/type-of-harm},
  note   = {Accessed on 2024-10-28}
}

@book{nagel1979mortal,
  title     = {Mortal questions},
  author    = {Nagel, Thomas},
  year      = {1979},
  publisher = {Cambridge University Press}
}

@article{quinn1989actions,
  title     = {Actions, intentions, and consequences: The doctrine of double effect},
  author    = {Quinn, Warren S},
  journal   = {Philosophy \& Public Affairs},
  pages     = {334--351},
  year      = {1989},
  publisher = {JSTOR}
}

@book{rochet2009there,
  title     = {Why are there so many banking crises? The politics and policy of bank regulation},
  author    = {Rochet, Jean-Charles},
  year      = {2009},
  publisher = {Princeton University Press}
}

@article{rudin2019stop,
  title     = {Stop explaining black box machine learning models for high stakes decisions and use interpretable models instead},
  author    = {Rudin, Cynthia},
  journal   = {Nature Machine Intelligence},
  volume    = {1},
  number    = {5},
  pages     = {206--215},
  year      = {2019},
  publisher = {Nature Publishing Group UK London}
}

@inproceedings{shelby2023sociotechnical,
  title     = {Sociotechnical Harms of Algorithmic Systems: Scoping a Taxonomy for Harm Reduction},
  author    = {Shelby, Renee and Rismani, Shalaleh and Henne, Kathryn and Moon, AJung and Rostamzadeh, Negar and Nicholas, Paul and Yilla-Akbari, N'Mah and Gallegos, Jess and Smart, Andrew and Garcia, Emilio and others},
  booktitle = {Proceedings of the 2023 AAAI/ACM Conference on AI, Ethics, and Society},
  pages     = {723--741},
  year      = {2023}
}

@book{shmueli2016practical,
  title     = {Practical time series forecasting with {R}: A hands-on guide},
  author    = {Shmueli, Galit and Lichtendahl Jr, Kenneth C},
  year      = {2016},
  publisher = {Axelrod Schnall Publishers}
}

@book{smiley1992moral,
  title     = {Moral responsibility and the boundaries of community: Power and accountability from a pragmatic point of view},
  author    = {Smiley, Marion},
  year      = {1992},
  publisher = {University of Chicago Press}
}

@article{smuha2021beyond,
  title   = {Beyond the individual: governing {AI}’s societal harm},
  author  = {Smuha, Nathalie A},
  journal = {Internet Policy Review},
  volume  = {10},
  number  = {3},
  year    = {2021}
}

@article{sornette2002predictability,
  title     = {Predictability of catastrophic events: Material rupture, earthquakes, turbulence, financial crashes, and human birth},
  author    = {Sornette, Didier},
  journal   = {Proceedings of the National Academy of Sciences},
  volume    = {99},
  number    = {suppl\_1},
  pages     = {2522--2529},
  year      = {2002},
  publisher = {National Acad Sciences}
}

@article{stilgoe2013developing,
  title   = {Developing a framework for responsible innovation},
  author  = {Stilgoe, JEZ and Macnaghten, P and Owen, R},
  journal = {Research Policy},
  volume  = {42},
  number  = {9},
  pages   = {1568--1580},
  year    = {2013}
}

@inproceedings{suresh2021framework,
  title     = {A framework for understanding sources of harm throughout the machine learning life cycle},
  author    = {Suresh, Harini and Guttag, John},
  booktitle = {Proceedings of the 1st ACM Conference on Equity and Access in Algorithms, Mechanisms, and Optimization},
  pages     = {1--9},
  number    = 17,
  series    = {EAAMO '21},
  year      = {2021},
  publisher = {Association for Computing Machinery},
  address   = {New York, USA}
}

@book{taylor1986respect,
  title     = {Respect for nature: A theory of environmental ethics},
  author    = {Taylor, Paul W},
  year      = {1986},
  publisher = {Princeton University Press}
}

@article{denton2021genealogy,
  title={On the genealogy of machine learning datasets: A critical history of ImageNet},
  author={Denton, Emily and Hanna, Alex and Amironesei, Razvan and Smart, Andrew and Nicole, Hilary},
  journal={Big Data \& Society},
  volume={8},
  number={2},
  pages={20539517211035955},
  year={2021},
  publisher={SAGE Publications Sage UK: London, England}
}

@book{onora2002autonomy,
  title={Autonomy and trust in bioethics},
  author={O’Neill, O},
  volume={37},
  year={2002},
  publisher={Cambridge University Press}
}

@article{villegas2023moral,
  title     = {Moral distance, {AI}, and the ethics of care},
  author    = {Villegas-Galaviz, Carolina and Martin, Kirsten},
  journal   = {AI \& Society},
  pages     = {1--12},
  year      = {2023},
  publisher = {Springer}
}

@inproceedings{weidinger2022taxonomy,
  title     = {Taxonomy of risks posed by language models},
  author    = {Weidinger, Laura and Uesato, Jonathan and Rauh, Maribeth and Griffin, Conor and Huang, Po-Sen and Mellor, John and Glaese, Amelia and Cheng, Myra and Balle, Borja and Kasirzadeh, Atoosa and others},
  booktitle = {Proceedings of the 2022 Conference on Fairness, Accountability, and Transparency},
  pages     = {214--229},
  year      = {2022},
  publisher = {Association for Computing Machinery}
}

@inproceedings{wieringa2020account,
  title     = {What to account for when accounting for algorithms: a systematic literature review on algorithmic accountability},
  author    = {Wieringa, Maranke},
  booktitle = {Proceedings of the 2020 Conference on Fairness, Accountability, and Transparency},
  pages     = {1--18},
  year      = {2020},
  publisher = {Association for Computing Machinery}
}

@article{hobday2019ethical,
  title={Ethical considerations and unanticipated consequences associated with ecological forecasting for marine resources},
  author={Hobday, Alistair J and Hartog, Jason R and Manderson, John P and Mills, Katherine E and Oliver, Matthew J and Pershing, Andrew J and Siedlecki, Samantha},
  journal={ICES Journal of Marine Science},
  volume={76},
  number={5},
  pages={1244--1256},
  year={2019},
  publisher={Oxford University Press}
}
